\documentclass[aps,showpacs]{revtex4}
\usepackage{graphicx,bm}
\usepackage{epsf}
\begin{document}
\title{The truncated Wigner method for Bose
condensed gases: limits
of validity and applications}
\author{Alice Sinatra}
\affiliation{Laboratoire Kastler Brossel,
Ecole Normale Sup\'erieure, UPMC and CNRS,
24 rue Lhomond, 75231 Paris Cedex 05, France}
\author{Carlos Lobo} 
\affiliation{Laboratoire Kastler Brossel,
Ecole Normale Sup\'erieure, UPMC and CNRS,
24 rue Lhomond, 75231 Paris Cedex 05, France}
\author{Yvan Castin}
\affiliation{Laboratoire Kastler Brossel,
Ecole Normale Sup\'erieure, UPMC and CNRS,
24 rue Lhomond, 75231 Paris Cedex 05, France}

\begin{abstract}

We study the truncated Wigner method applied to a weakly interacting
spinless Bose condensed gas which is perturbed away from thermal
equilibrium by a time-dependent external potential. The principle of the
method is to generate an ensemble of classical fields $\psi(r)$ which
samples the Wigner quasi-distribution function of the initial thermal
equilibrium density operator of the gas, and then to evolve each classical
field with the Gross-Pitaevskii equation. In the first part of the paper
we improve the sampling technique over our previous work [Jour. of Mod.
Opt. {\bf 47}, 2629-2644 (2000)] and we test its accuracy against the
exactly solvable model of the ideal Bose gas. In the second part of the
paper we investigate the conditions of validity of the truncated Wigner
method. For short evolution times it is known that the time-dependent
Bogoliubov approximation is valid for almost pure condensates. The
requirement that the truncated Wigner method reproduces the Bogoliubov
prediction leads to the constraint that the number of field modes in the
Wigner simulation must be smaller than the number of particles in the
gas. For longer evolution times the nonlinear dynamics of the noncondensed
modes of the field plays an important role. To demonstrate this we analyse
the case of a three dimensional spatially homogeneous Bose condensed gas
and we test the ability of the truncated Wigner method to correctly
reproduce the Beliaev-Landau damping of an excitation of the condensate.
We have identified the mechanism which limits the validity of the 
truncated Wigner method: the initial ensemble of classical fields, 
driven by the 
time-dependent Gross-Pitaevskii equation, thermalises to a classical field
distribution at a temperature $T_{\rm class}$ which is larger than the
initial temperature $T$ of the quantum gas. When $T_{\rm class}$
significantly exceeds $T$ a spurious damping is observed in the Wigner
simulation. This leads to the second validity condition for the truncated
Wigner method, $T_{\rm class} - T \ll T$, which requires that the maximum
energy $\epsilon_{\rm max}$ of the Bogoliubov modes in the simulation does
not exceed a few $k_B T$.

\end{abstract}


\maketitle
\section{Introduction}
In Ref. \cite{Drummond} the formalism of the
Wigner representation of the density operator, widely used in quantum
optics, was proposed as a possible way to study the time evolution of
Bose-Einstein condensates
in the truncated Wigner approximation \cite{wqo}.
Like other existing approximate
methods, such as the time-dependent Bogoliubov approach, 
it allows us to go beyond
the commonly used Gross-Pitaevskii equation, in which the interactions
between the condensate and the noncondensed atoms are neglected.
Our aim in this paper is to illustrate the advantages and the limits of the
truncated Wigner approach.

For reasons of clarity we will address two different situations
in two separate parts of the paper: (i) the case of a stationary Bose
condensed gas in
thermal equilibrium and (ii) a time-dependent case
when the gas is brought out of equilibrium by a known external
perturbation. 
Even if the stationary gas is the starting point for both
situations, the problems raised by the application of the 
Wigner method are of a different nature in the two cases.

(i) In the case of a Bose condensed gas in thermal equilibrium, the first
step is to calculate the Wigner quasi-distribution function associated
with the $N$-body density operator $\hat{\sigma}$, which is a functional
of a complex classical field $\psi(r)$.
We showed in \cite{PRL} that this is possible in the Bogoliubov 
approximation
when the noncondensed fraction of atoms is small. With such an
approximation, the Hamiltonian of the system is quadratic
in the noncondensed field and its Wigner functional
is a Gaussian. After that, we went through some more technical work
to calculate the Wigner functional of the whole matter
field including the condensate mode. 
In our derivation we made further approximations in addition to the
Bogoliubov approximation.  This introduces
some artifacts in the Wigner functional as far as the condensate mode
is concerned \cite{PRL}. These artifacts are, however, insignificant when
the number of thermally populated modes is much larger than one,
or $k_B T\gg\hbar\omega$ in an isotropic trap of harmonic frequency $\omega$,
so that the fluctuations in the number of condensate particles,
due to finite temperature, are much larger than one.
Once the Wigner functional for the Bose condensed gas in thermal
equilibrium is calculated, the goal is to be able to sample it
numerically in order to compute averages of observables and
probability distributions. In practice, this step consists in
generating a set of random atomic fields $\{\psi(r)\}$ according
to a probability distribution dictated by the Wigner functional.
We have now developed a more efficient algorithm to sample the Wigner
functional in the case of spatially inhomogeneous condensates in a
trapping potential than the one that we had presented in a previous paper
\cite{Kuhtai}, which we will explain here in detail. 
As far as the equilibrium Bose condensed gases are concerned, our method
in its regime of validity, is equivalent to the $U(1)$
symmetry-preserving
Bogoliubov approach of \cite{CastinDum,Gardiner}, up to
second
order in the small parameter of the theory, which is the square root
of the noncondensed fraction.
Compared with the traditional Bogoliubov approach,
our method presents, however, the practical advantage of avoiding the
direct diagonalisation of the Bogoliubov matrix, which is a heavy
numerical 
task in 2D and 3D in the absence of rotational symmetry. Moreover, due
to the stochastic formulation we adopt, 
our method gives us access to single realisations and to the probability
distribution of some observables such as the number of condensate
particles,
not easily accessible by the traditional Bogoliubov method.
We show some examples 
where we compare the probability distribution of the number of condensate
particles obtained with our method with an exact calculation in case of
the 
ideal Bose gas.

(ii) Let us now consider the situation of a
Bose condensed gas at thermal equilibrium which is brought out of equilibrium 
by a perturbation.  The initial Wigner functional then evolves in time according to a
kind of Fokker-Planck equation containing first and third order
derivatives with respect to the atomic field. Numerical simulation
of the exact evolution equation for the Wigner functional has
intrinsic difficulties, as one would expect, since it represents the exact
solution of the quantum many-body problem \cite{Iacopo}. We are less ambitious
here, and we rely on an approximation that consists in neglecting
the third order derivatives in the evolution equation. 
This is known as
the truncated Wigner approximation \cite{Drummond}.
For a
delta interaction potential between a finite number of low energy
modes of the atomic field, the third order derivatives are
expected to give a contribution which is smaller than that of the first
order derivatives when the occupation numbers of the modes are much larger
than unity. 
If we reason in terms of the stochastic
fields $\psi(r,t)$ which sample the Wigner distribution at time $t$, then the
truncated Wigner approximation corresponds to evolving the
initial set of stochastic fields according to the Gross-Pitaevskii
equation \cite{terme_omis}:
\begin{equation}
i\hbar \partial_t \psi = \left[-\frac{\hbar^2}{2m}\Delta
+U(r,t) + g |\psi|^2 \right]\psi,
\label{eq:GPE}
\end{equation}
where $r$ is the set of single particle spatial coordinates,
$m$ is the atom mass, $U$ is the trapping potential and
$g$ is the coupling constant originating from the effective
low energy interaction potential $V(r_1-r_2)=
g\delta(r_1-r_2)$ and proportional to the $s$-wave
scattering length $a$ of the true interaction potential,
$g=4\pi\hbar^2 a/m$. Here, the crucial difference
with respect to the usual Gross-Pitaevskii equation is that the field is
now
the whole matter field rather than the condensate field.

This procedure of evolving a set of random fields with the
Gross-Pitaevskii equation is known as the classical field
approximation, since equation (\ref{eq:GPE}) can be formally obtained
from the Heisenberg equation of motion for the atomic field operator
$\hat{\psi}$ by replacing the field operator by a c-number
field. The classical field approximation has already been used
in the Glauber-P representation to study the formation of the
condensate \cite{Kagan,Sachdev,Davis,Polonais,Burnett}.
We face here a different situation: we assume an initially existing
condensate and we use the Wigner representation, rather than the Glauber-P
representation. The Wigner representation is in fact known in
quantum optics to make the classical
field approximation more accurate than in the Glauber-P representation
because the ``right amount'' of quantum noise is contained
in the initial state \cite{qnoise}. For a single mode system 
with a Kerr type nonlinearity and an occupation
number $n$, the term neglected in the Wigner evolution equation is a third order derivative 
which is $1/n^2$ times smaller than the classical field term, 
whereas the term neglected in the Glauber-P evolution equation is a second order derivative,
which is only $1/n$ times smaller than the classical field term.
In the case of Bose-Einstein condensates however, we face a
highly multimode
problem and, therefore, the accuracy of the truncated Wigner approach
needs
to be
revisited. We approach this problem in the second part of the paper. 
The strategy we adopt is to compare
the predictions of the truncated Wigner method with existing
well-established
results: first with the time-dependent Bogoliubov approach and then with
the Landau-Beliaev damping of a collective excitation in a spatially
homogeneous condensate.

\section{Basic notations and assumptions}
\label{sec:basics}

\subsection{Model Hamiltonian on a discrete grid}
Let us express a simple quantity like
the mean atomic density using the Wigner representation:
\begin{equation}
\langle \hat{\psi}^\dagger(r)\hat{\psi}(r) \rangle =
	\langle {\psi}^*(r){\psi}(r) \rangle_W - 
	\frac{1}{2}\langle [\hat{\psi}(r),\hat{\psi}^\dagger(r)] \rangle,
\end{equation}
where $\langle\ldots\rangle_W$ represents the average over the Wigner
quasi-distribution function. This shows that the discretisation of the problem on a finite
grid is necessary to avoid
infinities: in the continuous version of the problem,
$[\hat{\psi}(r),\hat{\psi}^\dagger(r)]=\delta(0)=+\infty$. Physically
this divergence comes from the fact that,
in the Wigner point of view,
some noise is included in  each mode of 
the classical field $\psi$ to mimic quantum noise; this extra noise
adds up to infinity for a system with an infinite number of modes.
Therefore we use, from the beginning, a discrete formulation of our
problem
which will make it also suitable for numerical simulations. 

We consider a discrete spatial grid forming a box of length $L_\nu$
along the direction $\nu=x,y,z$ with an even number $n_\nu$ 
of equally spaced points. We denote ${\cal N}\equiv
\prod_\nu n_\nu$ the number of points on the grid, $V\equiv\prod_\nu
L_\nu$
the volume of the grid and $dV\equiv V/{\cal N}$ the volume of the unit
cell 
of the grid.
We take periodic boundary conditions in the box \cite{big_enough}.
We can then expand the field operator over plane waves
\begin{equation}
\hat{\psi}(r) =\sum_k \hat{a}_k \frac{1}{\sqrt{V}} e^{i k\cdot r}, 
\end{equation}
where $\hat{a}_k$ annihilates a particle of momentum $k$ and where
the components of $k$ are $k_\nu=2\pi j_\nu/L_\nu$ with the integers $j_\nu$
running from $-n_\nu/2$ to $n_\nu/2-1$.
We then have the inverse formula:
\begin{equation}
\hat{a}_k = dV \sum_{r} \frac{1}{\sqrt{V}} e^{-i k\cdot r} \hat{\psi}(r).
\end{equation} 
For each node $r_i$ on the spatial grid, we find the commutation relations
for the field operator:
\begin{equation}
[\hat{\psi}(r_i),\hat{\psi}^\dagger(r_j)] = \frac{1}{dV} \delta_{i,j}
\end{equation}
and the discretised model Hamiltonian that we use is:
\begin{equation}
\hat{H}=\sum_k \frac{\hbar^2 k^2}{2m} \hat{a}^\dagger_k \hat{a}_k + 
	dV \sum_{r} U(r) \hat{\psi}^\dagger(r) \hat{\psi}(r) +
	\frac{g}{2} dV \sum_{r}  
	\hat{\psi}^\dagger(r) \hat{\psi}^\dagger(r)
				\hat{\psi}(r) \hat{\psi}(r) \,.
\label{eq:discrHam}
\end{equation}
The first term in (\ref{eq:discrHam}) is the kinetic energy, which is easy
to 
calculate in the momentum representation.
In the position representation, the kinetic energy is a 
matrix that couples the ${\cal N}$ points of the grid. In the following
we
will write it as $p^2/2m$. The second term is the
trapping
potential. The last term represents the atomic 
interactions modeled by a discrete Kronecker $\delta$ potential
\begin{equation}
V(r_1-r_2) = \frac{g}{dV} \delta_{r_1,r_2},
\label{eq:modpot}
\end{equation}
with a coupling constant $g=4\pi\hbar^2 a/m$, where $a$ is the $s$-wave 
scattering length of the true interaction potential.

We indicate briefly some requirements for the discrete Hamiltonian
to be a good representation of reality. First, the spatial step
of the grid should be smaller than the macroscopic physical scales
of the problem:
\begin{equation}
dx_\nu \ll \xi \quad \mbox{and} \quad dx_\nu \ll \lambda,
\label{eq:dxassezpetit}
\end{equation}
where $\xi=1/\sqrt{8\pi\rho|a|}$ is the healing length for the maximal atomic
density $\rho$ and $\lambda=\sqrt{2\pi\hbar^2/m k_B T}$ is the thermal de Broglie
wavelength at temperature $T$.
Secondly, the spatial step of the grid should be larger than the
absolute value of the scattering length $a$:
\begin{equation}
dx_\nu \gg |a|,
\label{eq:noname}
\end{equation}
so that the scattering amplitude of the model potential (\ref{eq:modpot})
is indeed very close to $a$. Another way of saying this is that the
model potential (\ref{eq:modpot}) can be treated in the Born approximation
for the low energy waves. A more precise treatment, detailed in the appendix
\ref{appen:g0}, is to replace in (\ref{eq:modpot}) the coupling constant
$g$ by its bare value $g_0$ adjusted so that the scattering length of the
model potential on the grid is exactly equal to $a$.

\subsection{Wigner representation}

The Wigner quasi-distribution function associated with the $N$-body
density
operator $\hat{\sigma}$ 
is defined as the Fourier transform of the characteristic function $\chi$:
\begin{eqnarray} 
W(\psi) &\equiv& \int \prod_r \frac{d\mbox{Re}\,\gamma(r)\,
d\mbox{Im}\,\gamma(r)dV}{\pi^2}
\, \chi(\gamma)\, e^{ dV \sum_r \gamma^*(r) \psi(r)-\gamma(r)\psi^*(r)}
\label{eq:Wigner} \\
\chi(\gamma) &=&
\mbox{Tr}\left[\hat{\sigma} e^{dV\sum_r \gamma(r)
\hat{\psi}^{\dagger}(r)-\gamma^*(r)\hat{\psi}(r)} \right],
\label{eq:chi}
\end{eqnarray} 
where $\gamma(r)$ is a complex field on the spatial grid and
$\hat\sigma$ is the density operator of the system. With this
definition the Wigner function is normalised to unity:
\begin{equation}
\int \prod_r d\mbox{Re}\psi(r) d\mbox{Im}\psi(r)dV \, W(\psi)=1.
\end{equation}
We recall that the moments of the Wigner function correspond to totally
symmetrised quantum expectation values, i.e.,
\begin{equation}
\langle O_1 \ldots O_n\rangle_W =\frac{1}{n!}\sum_P
\mbox{Tr}\left[\hat{O}_{P(1)}\ldots\hat{O}_{P(n)}\hat\sigma
\right],
\end{equation}
where the sum is taken over all the permutations $P$ of $n$
objects, $O_k$ stands for $\psi$ or $\psi^*$ in some point of the
grid and $\hat{O}_k$ is the corresponding field operator.

The equation of motion for the density operator $\hat{\sigma}$
\begin{equation}
\frac{d}{dt}\hat{\sigma} = \frac{1}{i\hbar} [\hat{H},\hat{\sigma}]
\end{equation}
can be written exactly as the following equation of motion
for the Wigner distribution:
\begin{equation}
i\hbar\frac{\partial W}{\partial t}=\sum_r 
\frac{\partial}{\partial \psi(r)}(-f_\psi W)+\frac{g}{4 (dV)^2}
\frac{\partial^3 }{\partial^2 \psi(r) \partial \psi^*(r)} (\psi(r)W)
- \mbox{c.c.},
\end{equation}
with a drift term
\begin{equation}
f_\psi = \left[\frac{p^2}{2m} + U(r,t) + g \psi^*\psi 
-\frac{g}{dV}\right] \psi.
\label{eq:drift}
\end{equation}
The truncated Wigner approximation consists in neglecting the
cubic derivatives in the equation for $W$. The resulting equation
reduces to the drift term whose effect amounts to evolving
the field $\psi$ according to an equation which resembles the 
Gross-Pitaevskii 
equation (\ref{eq:GPE}).
The constant term $-g/dV$ inside the brackets of the above
equation can be eliminated by a redefinition of the global phase
of $\psi$, which has no physical consequence for observables
conserving the number of particles. 

\section{Sampling the Wigner functional for a Bose condensed gas
in thermal equilibrium}
\label{sec:equilibrium}

In \cite{PRL} we derive an expression of the Wigner functional for a
Bose condensed gas in thermal equilibrium in the frame of the $U(1)$
symmetry-preserving Bogoliubov approach \cite{CastinDum,Gardiner},
in which the gas has a fixed total number of particles equal to $N$.
We first introduce the approximate condensate wavefunction
$\phi(r)$, which is a solution of the time-independent Gross-Pitaevskii
equation:
\begin{equation}
\label{eq:tigpe}
H_{\rm gp} \phi \equiv
\left[\frac{p^2}{2m} + U(r,t=0) +
        N g |\phi|^2-\mu\right]\phi =0.
\end{equation}
We then split the classical field $\psi(r)$
into components orthogonal
and parallel to the condensate wavefunction $\phi(r)$:
\begin{eqnarray}
\label{eq:splite}
\psi(r) &=& a_\phi \phi(r) + \psi_\perp(r) \\
a_\phi &\equiv& dV \sum_{r} \phi^*({r}) \psi(r).
\end{eqnarray}
The Wigner functional provides us with the joint probability distributions
of
the transverse classical field $\psi_\perp(r)$, that we call
the noncondensed field, and of the
complex amplitude $a_\phi$.
Due to the $U(1)$ symmetry-preserving character of the theory, the
final Wigner functional is of the form \cite{PRL}
\begin{equation}
W(\psi) = \int \frac{d\theta}{2\pi} \; W_0(e^{-i\theta}\psi) .
\end{equation}
This means that one can sample the distribution $W(\psi)$
by (i) choosing a random field $\psi$ according to the distribution
$W_0(\psi)$, (ii) choosing a random global phase $\theta$ uniformly
distributed between 0 and $2\pi$, and (iii) forming the total atomic
field as $\psi_{\rm tot}(r) = e^{i\theta}\psi(r)$.
In practice, the global phase factor $e^{i\theta}$ is unimportant
to calculate the expectation
value of observables that conserve the number of particles. Since 
the other observables have a vanishing mean value,
we can limit ourselves to the sampling of the $\theta=0$
component of the Wigner functional, $W_0(\psi)$.

\subsection{Sampling the distribution of the noncondensed field}

The first step of the sampling procedure consists in generating a set
of noncondensed fields $\{\psi_\perp\}$ according to the probability
distribution
\begin{equation}
P(\psi_\perp) \propto \exp\left[
-dV \; (\psi_\perp^*,\psi_\perp) \cdot M
\left(
\begin{tabular}{c}
$\psi_\perp$ \\ 
$\psi_\perp^\ast$
\end{tabular}
\right)
\right]
\label{eq:Pperp},
\end{equation}
where we have collected all the components of $\psi_\perp$ and
$\psi_\perp^*$ in a single vector with $2{\cal N}$ components,
$M$ is the $2{\cal N}\times2{\cal N}$ matrix: 
\begin{equation}
\label{eq:M}
M = \eta \tanh \frac{\cal L}{2 k_B T}
\end{equation}
with
\begin{equation}
\eta = \left(
\begin{tabular}{rr}

1 & 0 \\
0 & $-1$
\end{tabular}
\right),
\end{equation}
and where ${\cal L}$ is a $2{\cal N}\times2{\cal N}$ matrix,
which is the discretised version of the Bogoliubov operator of
\cite{CastinDum}:
\begin{equation}
{\cal L} = \left(
\begin{tabular}{cc}
$H_{\rm gp} + N g {\cal Q} |\phi|^2 {\cal Q} $ & $N g {\cal Q} \phi^2 
	{\cal Q}^* $\\
$ - N g {\cal Q}^* \phi^{*2} {\cal Q} $ &  $-H_{\rm gp}^* - 
	N g {\cal Q}^* |\phi|^2 {\cal Q}^* $
\end{tabular}
\right).
\label{eq:calL}
\end{equation}
In this expression 
the ${\cal N}\times{\cal N}$ matrix ${\cal Q}$ projects
orthogonally to the condensate wavefunction $\phi$ in the discrete spatial
grid $\{r_i\}$ representation, 
\begin{equation}
{\cal Q}_{ij}=\delta_{ij}-dV\phi(r_i)\phi^*(r_j).
\label{eq:defQ}
\end{equation}
Note that the matrix $M$ can be shown to be Hermitian from the fact
that ${\cal L}^\dagger = \eta {\cal L}\eta$.

\subsubsection{Direct diagonalisation of ${\cal L}$}
If the eigenvectors of ${\cal L}$ are known, we
can use the following modal expansion:
\begin{equation}
\left(
\begin{tabular}{c}
$\psi_\perp$ \\ $\psi_\perp^*$
\end{tabular}
\right)
= \sum_k b_k
\left(
\begin{tabular}{c}
$u_k$ \\ $v_k$

\end{tabular}
\right)
+ b_k^*
\left(
\begin{tabular}{c}
$v_k^*$ \\ $u_k^*$
\end{tabular}
\right),
\end{equation}
where the sum is to be taken over all eigenmodes $(u_k,v_k)$
of $\cal L$ normalisable as
$\langle u_k|u_k\rangle-\langle v_k|v_k\rangle~=~1$, with corresponding
eigenvalues $\epsilon_k$. Since the condensate is assumed to be
in a thermodynamically stable or metastable state, all the $\epsilon_k$
are positive \cite{Houches}. The probability distribution
(\ref{eq:Pperp}) is then a simple product of Gaussian distributions
for the complex amplitudes $b_k$:
\begin{equation}
P_k(b_k) =\frac{2}{\pi}
\tanh\left(\frac{\epsilon_k}{2 k_B T}\right) \exp\left[
-2|b_k|^2\tanh\left(\frac{\epsilon_k}{2 k_B T}\right)\right].
\label{eq:distriG}
\end{equation}
Each Gaussian distribution is easily sampled numerically
\cite{NumRecGauss}. Note that, even at zero temperature, the Gaussian
distribution has a nonzero width: this is a signature of the extra
noise introduced in the Wigner representation to mimic quantum noise.

\subsubsection{Brownian motion simulation}
\label{sub:algorithm}

The sampling of the distribution (\ref{eq:Pperp}) can actually
be performed without  diagonalisation of $\cal L$ (an advantage
for spatially inhomogeneous Bose condensates in the absence of
rotational symmetry \cite{Kuhtai}) by means of a  
Brownian motion simulation for the noncondensed field: 
\begin{equation}
\label{eq:brown}
d
\left(\begin{tabular}{c}
$\psi_\perp$ \\ $\psi_\perp^*$
\end{tabular}
\right)
=
-\alpha \; dt
\left(\begin{tabular}{c}
$\psi_\perp$ \\ $\psi_\perp^*$
\end{tabular}
\right)
+Y
\left(\begin{tabular}{c}
$d\xi$ \\ $d\xi^*$
\end{tabular}
\right),
\end{equation}
where the field $d\xi$ is the noise term.
The time $t$ here is a purely fictitious time
with no physical meaning and will be taken to be
dimensionless. On our discrete grid, $\psi_\perp$
is a vector with $\cal N$ components, $d\xi$ is a Gaussian
random vector of $\cal N$ components with zero mean and a covariance
matrix $\langle d\xi_i d\xi_j^*\rangle$ equal 
to $(2 dt/dV)\delta_{i,j}$, while $\alpha, Y$ are
$2{\cal N}\times 2{\cal N}$ matrices.
To ensure that the Brownian motion relaxes towards the correct probability
distribution (\ref{eq:Pperp}) we require that the drift matrix $\alpha$
and the diffusion matrix $D\equiv Y (Y^\dagger)$ satisfy a generalised
Einstein's relation \cite{Kuhtai}:
\begin{equation}
\label{eq:Einstein}
D^{-1}\alpha =\alpha^\dagger D^{-1}= 2 M,
\end{equation}
where $M$ is the matrix (\ref{eq:M}).
There is, of course, no unique choice for $\alpha$ and $Y$. With respect
to our previous work \cite{Kuhtai}, we have largely improved
the efficiency of our simulation by a different choice of $\alpha, Y$
and by the use of a second order integration scheme of the
stochastic differential equation (\ref{eq:brown}), more efficient
than the usual first order Euler's scheme. In the appendix
\ref{app:stoch} we give
a detailed description of these improvements, useful to the reader
who is interested
in implementing the numerical algorithm.

\subsection{Sampling the condensate amplitude}
We now have to sample the condensate amplitude $a_\phi$
from the Wigner functional $W_0$.
This amplitude turns out to be real, and can
be written as
\begin{equation}
a_\phi =\sqrt{N_0} \hspace{1cm} 
\mbox{where} \hspace{1cm} N_0=a_\phi^* a_\phi \,.
\end{equation}
Since we already know how to generate the noncondensed part of the field
${\psi_\perp}$, we have to sample the conditional distribution 
$P(N_0|\psi_\perp)$. 

Due to a first approximation that we have performed in \cite{PRL}, 
which consists in treating ``classically'' the condensate mode and 
neglecting its quantum fluctuations in the limit of a 
large number of condensate particles, the probability 
distribution $P(N_0)$, that we will obtain by averaging $P(N_0|\psi_\perp)$
over the stochastic realisations of the noncondensed field $\psi_\perp$,
actually coincides with the 
probability distribution of the number of condensed particles 
$\hat{a}_\phi^\dagger  \hat{a}_\phi$ so that within this approximation
we have:
\begin{eqnarray}
\langle N_0 \rangle&=&\langle \hat{a}_\phi^\dagger \hat{a}_\phi \rangle, \\
\mbox{Var}(N_0)&=&\mbox{Var}(\hat{a}_\phi^\dagger \hat{a}_\phi),...
\end{eqnarray} 
Note that this should not be the case
for the exact Wigner distribution as, e.g., the average $\langle N_0
\rangle$ 
should be equal to $\langle \hat{a}_\phi^\dagger  \hat{a}_\phi \rangle + 1/2$
and the variance of $N_0$ should exceed the variance of 
$\hat{a}_\phi^\dagger  \hat{a}_\phi$ by $1/4$.

We show in \cite{PRL} that, 
when the number of thermally populated modes is much larger than one,
the width in $N_0$ of the conditional distribution $P(N_0|\psi_\perp)$ 
is much narrower than the width of the distribution $P(N_0)$, so that
we can replace the distribution $P(N_0|\psi_\perp)$ by a delta function
centered on its mean value.
With this second, more severe, approximation we get for the sampling:
\begin{equation}
N_0 \simeq \mbox{Mean}( N_0| \psi_{\perp})= 
	C - \frac{1}{2} dV (\psi_\perp^*,\psi_\perp)\cdot
	\left[\mbox{Id}-M^2 \right] \pmatrix{\psi_\perp \cr \psi_\perp^* \cr}
\label{eq:meancondN0},
\end{equation}
where the constant $C$ is finite only in the discretised version and is
given by
\begin{equation}
C= N-\frac{1}{4}\mbox{Tr}\,M +\frac{1}{2}\mbox{Tr}\,{\cal Q}.
\end{equation}
Here, the trace of the projector ${\cal Q}$ is simply the number of modes 
in the simulation minus one.

The second approximation (\ref{eq:meancondN0}) does not introduce 
errors in the average $\langle N_0 \rangle$. 
We are able to verify a posteriori that the error introduced in the
variance
$\langle N_0^2 \rangle-\langle N_0 \rangle^2$ is small in the following
way: on one hand we calculate the variance of $N_0$ (Var$(N_0)$),
by using (\ref{eq:meancondN0}). On the other hand we calculate the
variance Var$(\hat{\delta N})$ of the number of noncondensed particles
by using directly the ensemble of noncondensed
fields $\{\psi_\perp\}$. Since the total number of particles is fixed
one should have 
$\mbox{Var}(N_0)=\mbox{Var}(\hat{a}_\phi^\dagger\hat{a}_\phi)=
 \mbox{Var}(\hat{\psi}_\perp^\dagger \hat{\psi}_\perp)$, and deviation from
this identity gives us the error of Var$(N_0)$.

We are now ready to form the total field:
\begin{equation}
\psi(r)=\sqrt{N_0}\left(\phi(r)+\frac{\phi_\perp^{(2)}(r)}{N}\right)+
\psi_\perp(r).
\label{eq:initial}
\end{equation}
The function $\phi_\perp^{(2)}$ is a correction to the condensate wavefunction including
the condensate depletion neglected in the Gross-Pitaevskii equation (\ref{eq:tigpe})
and the mean field effect of the noncondensed particles.
This correction  can be calculated from the ensemble of noncondensed
fields
$\{\psi_\perp\}$ as explained in \cite{Kuhtai}. As we will see in
section \ref{sub:Bogol}
its contribution to the one-body density matrix is of the same
order
as that of $\psi_\perp$ and therefore has to be included.

\subsection{Tests and applications: Distribution of the number of condensate
particles}
\label{sub:tests}

We can use the sampling procedure described above
to calculate some equilibrium properties of the Bose condensed gas.
Recently, the variance of the number 
of particles in the condensate has drawn
increasing attention \cite{Wilkens,Stringari,Scully}.
In our case we have access to the whole probability distribution
for $N_0$ by applying equation (\ref{eq:meancondN0}) to the ensemble 
of stochastic noncondensed fields $\{\psi_\perp\}$.

\subsubsection{Ideal Bose gas}
\begin{figure}[htb]
\centerline{\epsfxsize=7cm \epsfbox{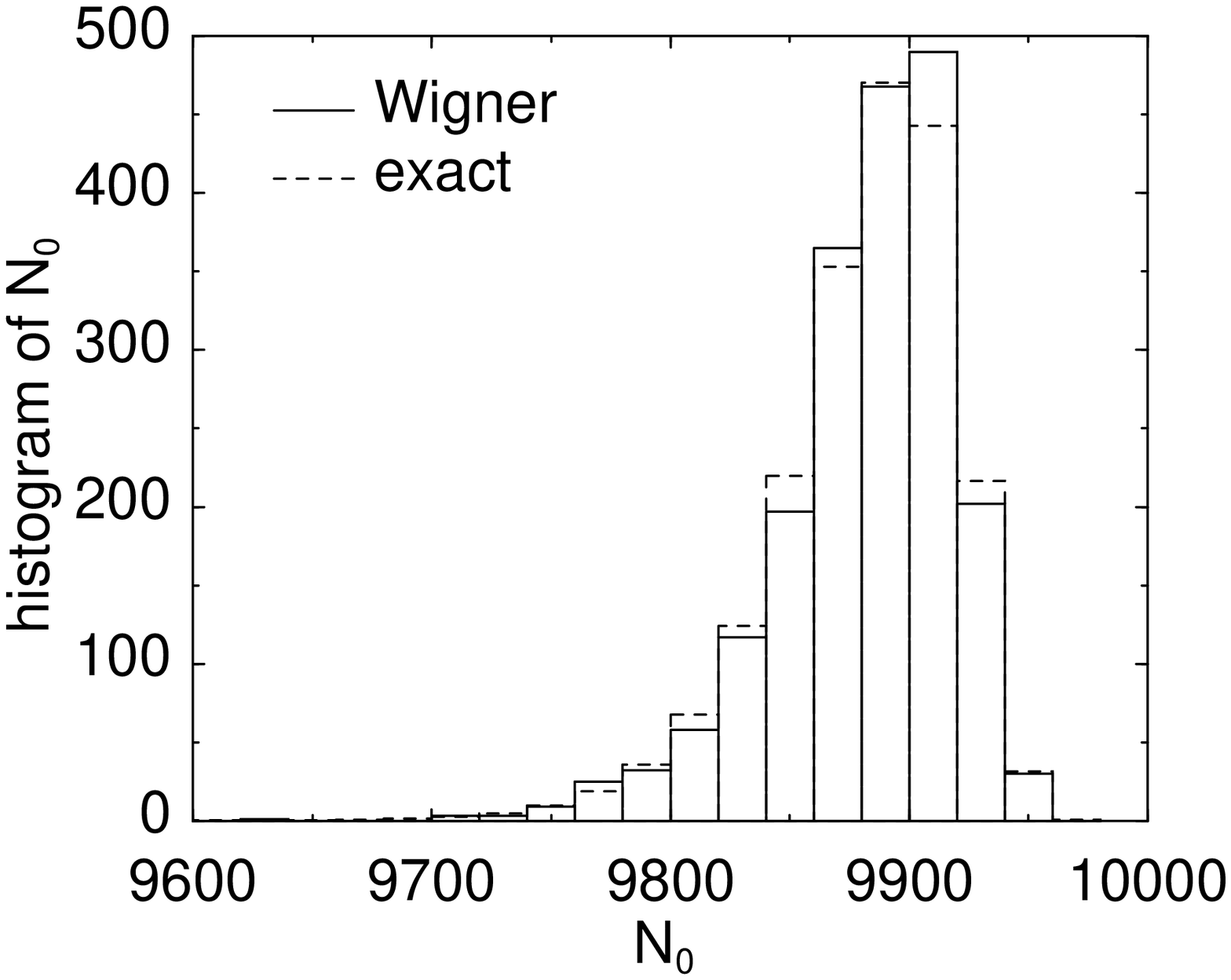} \hfill
    \epsfxsize=7cm \epsfbox{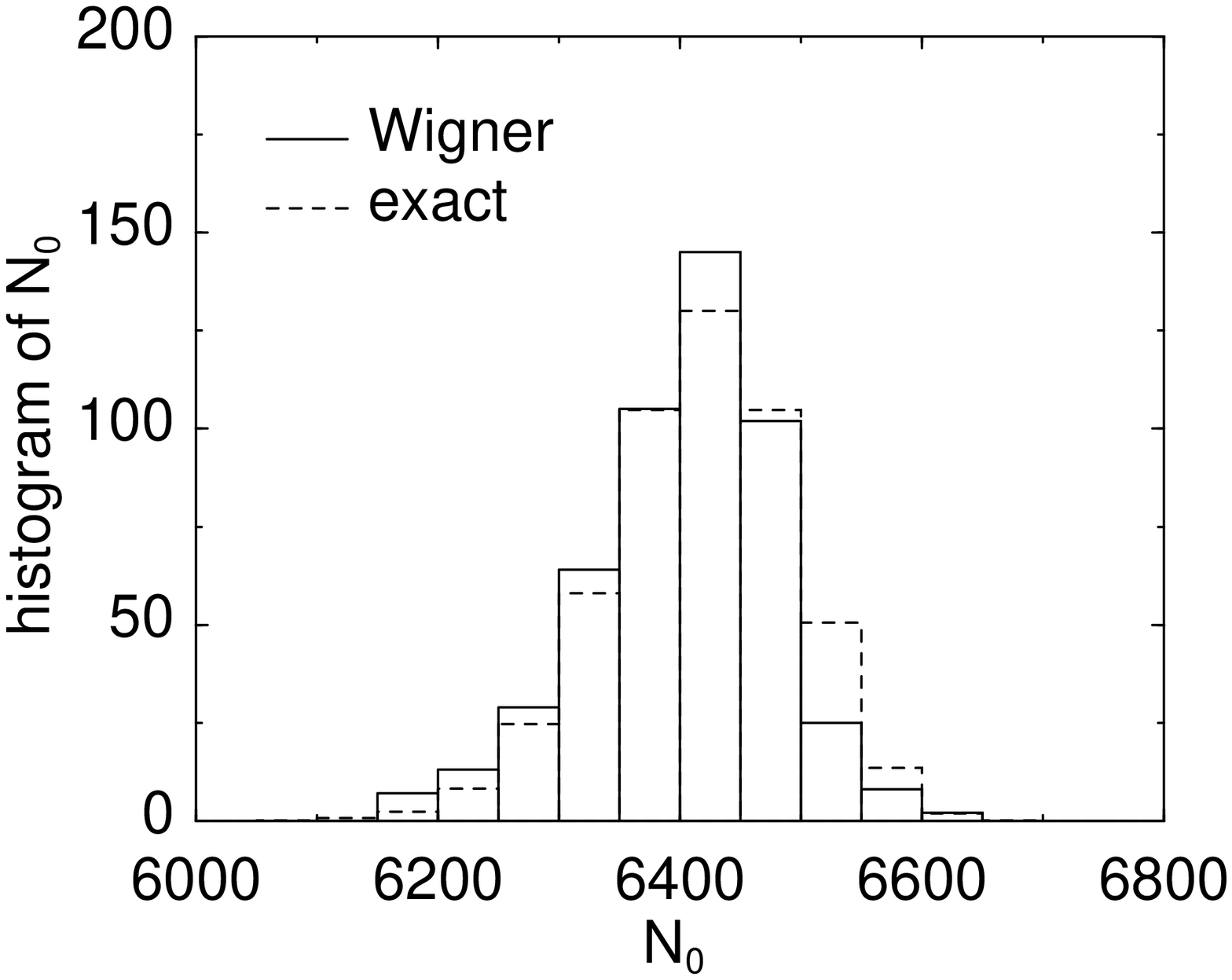}} 
\caption{
Probability distribution in the canonical
ensemble
of the number of condensate 
particles for the ideal Bose gas in thermal equilibrium
in an isotropic  harmonic potential $U(r) =
\frac{1}{2}m\omega^2 r^2$. (a) In a 1D model
for $k_B T=30 \hbar \omega$,  and $N=10\,000$.
For the Wigner simulation
2000 realisations have been performed on a grid with 128 points.
For the exact Bogoliubov rejection method described in the end of 
this subsection on the ideal gas, 
400 000 realisations have been performed so that the statistical error is less
than one per cent for the most populated channels of the histogram.
 (b) In a 2D model for
$k_B T=30 \hbar \omega$,  and $N=8\,000$.
For the Wigner simulation 
500 realisations have been performed on a grid with $128\times
128$ points. For  the exact sampling 100 000 realisations have been performed.}
\label{fig:ideal}
\end{figure}

As a test we check our probability distribution  for the number of condensate
particles against the exact one for the ideal Bose gas $(g=0)$ in one and two
dimensions. The results are in figure \ref{fig:ideal}.

The distributions of the number of condensed particles
$N_0$ are clearly not Gaussian. To characterise them, besides
the mean and the variance of $N_0$
one can introduce the skewness defined as:
\begin{equation}
\mbox{skew}(N_0)=\frac{\langle (N_0- \langle N_0 \rangle)^3 \rangle}
		{(\langle N_0^2 \rangle-\langle N_0 \rangle^2)^{3/2}}.
\end{equation}
For the parameters of figure \ref{fig:ideal} we give the mean, the 
standard deviation and the skewness
of $N_0$ obtained from the simulation, together
with their exact values:

\bigskip
\begin{center}
\begin{tabular}{|c||c|c||c|c|}
\hline 
 & 1D simulation & 1D exact 
	& 2D simulation & 2D exact  \\
\hline
$\langle N_0\rangle$  & 9882. & 9880. & 6403. &  6415.  \\
\hline
$\Delta N_0$ & 37.5 & 38.3 & 75.9 & 77.1 \\
\hline
skew$(N_0)$ & $-1.20$ & $-$1.16 & $-$0.40 & $-$0.334 \\
\hline
\end{tabular}
\end{center}

\bigskip
In what follows we explain in some detail how the exact 
probability distribution for the ideal Bose gas is obtained. 
Let $\hat{\sigma}$ be the density operator
for the ideal Bose gas in the canonical ensemble: 
\begin{equation}
\hat{\sigma}=\frac{1}{Z} \, e^{-\beta \hat{H}} \, p_N .
\end{equation}
The operator $p_N$ projects onto the subspace with $N$ particles, and 
$\hat{H}=\sum_k \epsilon_k \hat{a}_k^\dagger  \hat{a}_k$ is written in the 
eigenbasis of the trapping potential.
In the spirit of the number conserving Bogoliubov method, 
we eliminate the
condensate mode by writing
\begin{equation}
\hat{a}_0^\dagger \hat{a}_0 = 
	\hat{N} - \sum_{k \neq 0} \hat{a}_k^\dagger  \hat{a}_k \,.
\label{eq:condmode}
\end{equation}
Since the total number of particles is fixed we
can replace the operator
$\hat{N}$ by the c-number $N$ in (\ref{eq:condmode}). 
Furthermore we establish a one to one correspondence 
between (i) each configuration of excited modes $\{n_k,k>0\}$
having a number of excited particles $N^\prime=\sum_k n_k$ lower than $N$
and (ii) each configuration of the whole system with $n_k$ particles
in excited mode $k$ and $N-N'$ particles in the condensate.
We then obviously have to reject the configurations of excited modes
for which the number of particles in the excited states $N'$
is larger than $N$. This amounts to reformulating 
the effect of the projector $p_N$ in terms of an Heaviside function 
$Y$.  We then rewrite $\hat{\sigma}$ as:
\begin{equation}
\label{eq:demon}
\hat{\sigma}=\frac{1}{Z} \, e^{-\beta \epsilon_0 {N}} \,
	e^{-\beta \sum_{k \neq 0} (\epsilon_k-\epsilon_0) 
	\hat{a}_k^\dagger  \hat{a}_k }\, 
 Y\left( N-\sum_{k \neq 0} \hat{a}_k^\dagger  \hat{a}_k \right).
\end{equation}
For the sampling procedure we use a rejection method
i.e. we sample the probability distribution of the number of particles $n_k$
in each mode $k\neq 0$ without
the constraint imposed by the Heaviside function
and we reject configurations with
$N^\prime > N$. In this scheme we have to generate the $n_k$,
$k=1,\ldots,{\cal N}$,
according to the probability distribution 
\begin{equation} 
p_k(n_k)=\lambda_k^{n_k} (1-\lambda_k)\quad\mbox{with}\quad
\lambda_k=e^{-\beta  (\epsilon_k-\epsilon_0)}.
\end{equation} 
For each $k$ we proceed as follows: in a loop over $n_k$ starting from $0$
we generate a random number $\epsilon$ uniformly distributed in 
the interval $[0,1]$ and we compare it
with $\lambda_k$: if $\epsilon < \lambda_k$, we proceed with the next
step of the loop,
otherwise we exit from the loop and the current
value of $n_k$ is returned.

The calculation can also be done in the Bogoliubov approximation,
that is by neglecting the Heaviside function in (\ref{eq:demon}).
For the parameters of figure \ref{fig:ideal} this is actually
an excellent approximation, as the mean population of the
condensate mode is much larger than its standard deviation, and
the corresponding approximate results are in practice indistinguishable
from the exact ones.
The predictions of this Bogoliubov approximation for the first three
moments of $N_0$ involve a sum over all the excited modes of the
trapping potential:
\begin{eqnarray}
\langle N_0\rangle &=& N  -\sum_{k\neq 0} \bar{n}_k\nonumber \\
\mbox{Var}(N_0) &=& \sum_{k\neq 0} \bar{n}_k(1+\bar{n}_k)\nonumber \\
\langle \left(N_0-\langle N_0\rangle\right)^3\rangle &=&
\sum_{k\neq 0} 2\bar{n}_k^3+3\bar{n}_k^2+\bar{n}_k
\label{eq:sommes}
\end{eqnarray}
where $\bar{n}_k=1/(\exp(\beta(\epsilon_k-\epsilon_0))-1)$ is the
mean occupation number of the mode $k$.
In the limit $k_B T \gg \hbar \omega$ for an isotropic harmonic trap 
an analytical calculation, detailed in the appendix \ref{appendix:moments},
shows that the skewness tends to a constant in 1D, tends to
zero logarithmically in 2D
and tends to zero polynomially in 3D \cite{bizarre}:
\begin{eqnarray}
\mbox{skew}_{\rm 1D}(N_0) &\simeq & -\frac{2\zeta(3)}{\zeta(2)^{3/2}}
=-1.139547\ldots\nonumber \\
\mbox{skew}_{\rm 2D}(N_0) &\simeq & -\frac{2(\zeta(2)+\zeta(3))}{
(\log(k_BT/\hbar\omega) +1+\gamma+\zeta(2))^{3/2}}\nonumber \\
\mbox{skew}_{\rm 3D}(N_0) &\simeq & -\frac{\log(k_B T/\hbar\omega)+\gamma
+\frac{3}{2}+3\zeta(2)+2\zeta(3)}
{ (k_B T/\hbar\omega)^{3/2} \{\zeta(2)+(3\hbar\omega/2k_B T)
[\log(k_B T/\hbar\omega)+\gamma+1-\zeta(2)/3]\}^{3/2}
}
\label{eq:asympt}
\end{eqnarray}
where $\zeta$ is the Riemann Zeta function and $\gamma=0.57721\ldots$
is Euler's constant.

\subsubsection{Interacting case}
\begin{figure}[htb]
\centerline{\epsfxsize=7cm \epsfbox{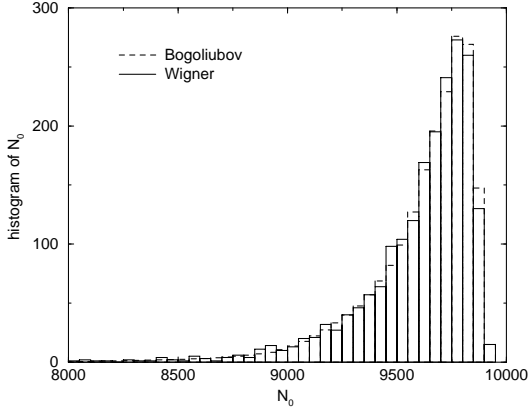}}
\caption{
Probability distribution of the number of condensate 
particles in the canonical ensemble for a 1D interacting Bose gas in
thermal equilibrium in a harmonic trap $U(x)=\protect\frac{1}{2}m\omega^2x^2$,
with $k_B T=30 \hbar \omega$, $\mu=14.1\hbar \omega$ and $N=10\,000$,
corresponding to a coupling constant $g=0.01 \hbar\omega (\hbar/m\omega)^{1/2}$.
The results have been obtained with the Wigner method
using 2000 realisations on a grid with 128 points. The dashed line is the 
histogram of the probability distribution of $N_0$ in the Bogoliubov
approximation generated using the same 2000 realisations, obtained with a method described in
\protect\cite{future}.}
\label{fig:inter} \end{figure}

As an example we show in figure \ref{fig:inter} the probability 
distribution for the number of condensate particles in the interacting case 
to demonstrate that the large skewness of $N_0$ in 1D
can even be enhanced in presence of interaction:  the skewness of $N_0$ in
figure \ref{fig:inter} is equal to $-2.3$. We have been able
\cite{future} to calculate $P(N_0)$ in the Bogoliubov
approximation in the interacting case starting from the sampling of the 
Wigner distribution of the noncondensed field (\ref{eq:Pperp}). We
compare the results with the Wigner approach in the same figure. As
expected the agreement is excellent in the regime $k_B T = 30 \hbar \omega
\gg \hbar \omega$.

\section{The truncated Wigner method for a time-dependent Bose condensed
gas }
\label{sec:validity}

In this section we investigate the conditions of validity of the
truncated Wigner approach for time-dependent Bose-Einstein condensates.
The strategy that we adopt is to compare the predictions of the
truncated Wigner approach to well-established theories: 
the time-dependent Bogoliubov approach in section \ref{sub:Bogol}
and the Landau-Beliaev damping of a collective excitation in a spatially
homogeneous condensate, in section \ref{sub:Landau}.

\subsection{The truncated Wigner method vs the time-dependent Bogoliubov
method}
\label{sub:Bogol}
In this section we investigate analytically the equivalence between the
time-dependent Bogoliubov approach of \cite{CastinDum} and the truncated 
Wigner method in the limit in which the noncondensed fraction is small.

We begin by sketching the number conserving Bogoliubov method
of Ref. \cite{CastinDum}.
We split the atomic field operator into components
parallel and orthogonal to the
exact time-dependent condensate wavefunction $\phi_{\rm ex}$ \cite{PENROSE} (omitting
for simplicity the time label for the field operators and for the
condensate wavefunction):
\begin{equation}
\hat{\psi}(r)=\hat{a}_{\phi_{\rm ex}} \phi_{\rm ex}(r) + \hat{\psi}_\perp(r)
\end{equation}
and we consider the limit
\begin{equation}
N \rightarrow \infty \hspace{1cm} N\,g = \mbox{constant}  
\hspace{1cm} T=\mbox{constant} \hspace{1cm} {\cal N}=\mbox{constant} .
\label{eq:limit}
\end{equation}
In \cite{CastinDum} one performs a formal systematic expansion in powers
of $1/\sqrt{N}$ of the exact condensate wavefunction
$\phi_{\rm ex}$
\begin{equation}
\phi_{\rm ex}(r)=\phi(r) + \frac{\phi^{(1)}(r)}{\sqrt{N}} +
\frac{\phi^{(2)}(r)}{N} + \ldots
\end{equation}
and of the noncondensed field 
\begin{equation}
\hat{\Lambda}_{\rm ex}(r) \equiv
\frac{1}{\sqrt{N}}\hat{a}_{\phi_{\rm ex}}^\dagger
				\psi_\perp(r) =
\hat{\Lambda}(r)+\frac{1}{\sqrt{N}}\hat{\Lambda}^{(1)}(r) + \ldots .
\label{eq:defLambdaex}  
\end{equation}
Note that in the lowest order approximation to $\hat{\Lambda}_{\rm ex}$ 
the exact condensate wavefunction $\phi_{\rm ex}$ is replaced by 
the solution $\phi$ of the time-dependent Gross-Pitaevskii equation
\begin{equation}
\label{eq:gpe_phi}
i\hbar\partial_t\phi = \left[p^2/2m + U(r,t)
+N g |\phi|^2\right]\phi
\end{equation}
and $\hat{a}_{\phi}/\sqrt{N}$ is replaced by the phase operator
$\hat{A}_{\phi}=\hat{a}_\phi{(\hat{a}_\phi^\dagger\hat{a}_\phi)}^{-1/2}$
so that
\begin{equation}
\hat{\Lambda}(r) = \frac{1}{\sqrt{\hat{a}_\phi^\dagger\hat{a}_\phi}}
\hat{a}_\phi^\dagger \left[ \hat{\psi}(r)-\phi(r)\hat{a}_\phi \right]
\end{equation}
and $\hat{\Lambda}(r)$ satisfies bosonic commutation relations
\begin{equation}
[\hat{\Lambda}(r),\hat{\Lambda}^\dagger(s)] = \frac{1}{dV}
{\cal Q}_{r,s}
\end{equation}
where the matrix
${\cal Q}_{r,s}=\delta_{r,s}-dV\phi(r)\phi^*(s)$ projects orthogonally to $\phi$.
To the first two leading orders in $1/\sqrt{N}$ one obtains 
an approximate form of the one-body density matrix:
\begin{eqnarray}
\langle r|\rho|s \rangle \equiv
        \langle \hat{\psi}^\dagger(s) \hat{\psi}(r) \rangle &=&
        (N-\langle \hat{\delta N} \rangle ) \phi(r)\phi^*(s) \nonumber \\
&& \nonumber \\
&+& \langle \hat{\Lambda}^\dagger(s) \hat{\Lambda}(r)\rangle \nonumber \\
&& \nonumber \\
&+& \phi^*(s)\phi_\perp^{(2)}(r) + \phi(r)\phi_\perp^{(2)*}(s) \nonumber \\
&& \nonumber \\
	&+& O(\frac{1}{\sqrt{N}}).
\label{eq:rho1quant}
\end{eqnarray}
We call the first term ``parallel-parallel'' because it originates
from the product of two parts of the field both parallel to the
condensate wavefunction; it describes the physics of a pure condensate with
$N-\langle \delta\hat{N} \rangle$ particles.
The second term, which we call ``orthogonal-orthogonal'' because
$\hat{\Lambda}$ is
orthogonal to $\phi$, describes the noncondensed
particles in the Bogoliubov approximation. The third term, called
``orthogonal-parallel'', describes corrections to the Gross-Pitaevskii
condensate wavefunction due to the presence of noncondensed particles
\cite{CastinDum}. In (\ref{eq:rho1quant}) $\langle \delta\hat{N} \rangle $ 
is the average number of noncondensed particles in the Bogoliubov
approximation:
\begin{equation}
\langle \delta\hat{N} \rangle =
    \sum_r \, dV \, \langle \hat{\Lambda}^\dagger(r) \hat{\Lambda}(r)\rangle.
\label{eq:dNBogol}
\end{equation}
The evolution equations for $\hat{\Lambda}$ and $\phi^{(2)}_\perp$
are given in appendix \ref{app:BogolCD}.

Having described the Bogoliubov method,
let us now consider the truncated Wigner approach in the limit 
(\ref{eq:limit}). We expand the classical field in powers of $1/\sqrt{N}$:
\begin{equation}
\psi=\sqrt{N} \psi^{(0)}+ \psi^{(1)} + \frac{1}{\sqrt{N}} \psi^{(2)} + \ldots
\label{eq:expansion}
\end{equation}
where the $\psi^{(j)}$ are of the order of unity.
We immediately note that the leading term of this expansion corresponds
to a pure condensate with $N$ particles so that $\psi^{(0)}$ is
simply the solution of the time-dependent Gross-Pitaevskii equation
(\ref{eq:gpe_phi}), $\psi^{(0)}=\phi$. This physically clear fact
will be checked explicitly in what follows.
In the initial thermal equilibrium state at time $t=0$  
we expand (\ref{eq:initial}) in powers of $1/\sqrt{N}$:
\begin{equation}
\sqrt{N_0}\equiv \sqrt{N-\delta N} = \sqrt{N} -
                \frac{1}{2}\frac{\delta N}{ \sqrt{N} } + \ldots 
\end{equation}
so that we can identify explicitly:
\begin{eqnarray}
\psi^{(0)}(t=0)&=&\phi \label{eq:ordrezero}\\
\psi^{(1)}(t=0)&=&\psi_\perp \label{eq:ordreun} \\
\psi^{(2)}(t=0)&=&-\frac{\delta N}{2} \phi + \phi_\perp^{(2)}.
\label{eq:ordredeux}
\end{eqnarray}
Following the same procedure as in the quantum case, we split
each term of the expansion into a component along $\phi$
and a component orthogonal to $\phi$:
\begin{equation}
\psi^{(j)}(r)=\xi^{(j)}\phi(r)+\psi_\perp^{(j)}.
\label{eq:split}
\end{equation}
We calculate now the one-body density matrix ${\rho}$.
Since we are using the Wigner representation for the atomic field
on a finite spatial grid we have:
\begin{equation}
\langle r |\hat{\rho}| s \rangle = \langle {\psi}^*(s) \psi(r) \rangle
        - \frac{1}{2 dV} \delta_{r,s}
\label{eq:rho0wig}
\end{equation}
where $dV$ is the unit cell volume of the spatial grid and $\delta_{r,s}$
is a Kronecker $\delta$. Note that to simplify the notation we have omitted
the subscript $W$ on the right hand side of the equation since
the quantum and Wigner averages can be readily distinguished by the
hats on the operators.
We insert the expansions (\ref{eq:expansion}) and (\ref{eq:split}) into
(\ref{eq:rho0wig}) and we use the fact that $\psi^{(0)}=\phi$
to obtain:
\begin{eqnarray}
\langle r|\hat{\rho}|s \rangle_{TW} &=& 
  \phi^*(s)\phi(r) \left[N+ 
   \sqrt{N} \langle \xi^{(1)}+\xi^{(1)*} \rangle   +
 \langle |\xi^{(1)}|^2  \rangle +  \langle \xi^{(2)}+\xi^{(2)*} \rangle 
  - \frac{1}{2}      \right]
\nonumber \\
&& \nonumber \\
&+&  \langle \psi^{(1)*}_\perp(s) \psi^{(1)}_\perp(r) \rangle 
- \frac{1}{2 dV} {\cal Q}_{r,s} \nonumber \\
&& \nonumber \\
&+& \; { \phi^*(s)} [ { \sqrt{N}}  \langle \psi^{(1)}_\perp(r)\rangle  + 
         \langle \xi^{(1)*} \psi^{(1)}_\perp(r) \rangle +
        \langle \psi^{(2)}_\perp(r) \rangle ]
	+ { \{r \leftrightarrow s \}^*} \nonumber \\
&&	\nonumber \\
&+& O\left(\frac{1}{\sqrt{N}}\right) 
\label{eq:rho1wig}
\end{eqnarray}
where we have collected the terms ``parallel-parallel'' in the first line,
the terms ``orthogonal-orthogonal'' in the second line and the terms
``orthogonal-parallel'' in the third line, and where the matrix
${\cal Q}_{r,s}/dV = \delta_{r,s}/dV - \phi(r)\phi^*(s)$ is the discrete
version of the projector $Q=1-|\phi\rangle\langle\phi|$.
As we show in appendix \ref{app:Bogol}, by using the evolution equation
of the field (\ref{eq:GPE}) and the initial conditions 
(\ref{eq:ordrezero}), (\ref{eq:ordreun}) and(\ref{eq:ordredeux})
the following identities hold at all times:
\begin{eqnarray}
\psi^{(0)}&=&\phi \label{eq:id_psi0} \\
   \sqrt{N} \langle \xi^{(1)}+\xi^{(1)*} \rangle   +
 \langle |\xi^{(1)}|^2  \rangle +  \langle \xi^{(2)}+\xi^{(2)*} \rangle
 &=& - \langle \delta \hat{N} \rangle \label{eq:id_deltaN} \\ 
\langle \psi^{(1)*}_\perp(s) \psi^{(1)}_\perp(r) \rangle
- \frac{1}{2 dV} {\cal Q}_{r,s} &=&
\langle \hat{\Lambda}^\dagger(s) \hat{\Lambda}(r)\rangle
\label{eq:lin} \\
  { \sqrt{N}}  \langle \psi^{(1)}_\perp(r)\rangle  + 
         \langle \xi^{(1)*} \psi^{(1)}_\perp(r) \rangle +
        \langle \psi^{(2)}_\perp(r) \rangle &=& \phi^{(2)}_\perp(r) .
\label{eq:id_phi2}
\end{eqnarray}
As we have already mentioned the first identity
(\ref{eq:id_psi0}) reflects the fact that at zero
order in the expansion we have a pure condensate with $N$ particles
evolving according
to the time-dependent Gross-Pitaevskii equation.
At time $t=0$ the three other identities are easily established since we
have simply $\langle\psi_\perp^{(1)}\rangle=0$, $\xi^{(1)}=0$ and
$\xi^{(2)}=- \delta N/2$.
At later times the mean value $\langle\psi_\perp^{(1)}\rangle$ remains
equal to zero while $\xi^{(1)}$ develops a nonzero imaginary part
corresponding to phase change of $\psi$ in the mode $\phi$ due to the
interaction with the
noncondensed particles
\begin{equation}
\psi=\sqrt{N} \phi + \xi^{(1)} \phi + \ldots \simeq \sqrt{N}
        e^{\xi^{(1)}/\sqrt{N}} \phi + \ldots
\end{equation}
After averaging over all stochastic realisations, this random phase change
contributes to the condensate depletion in
(\ref{eq:id_deltaN}) and to the correction $\phi^{(2)}$ to the condensate
wavefunction in (\ref{eq:id_phi2}) \cite{enfin}.
As a consequence of the purely imaginary character of $\xi^{(1)}$
the quantity proportional to $\sqrt{N}$ in (\ref{eq:id_deltaN}) vanishes.
The identity (\ref{eq:lin}) reflects the fact that in the linearised
regime
quantum fluctuations (here $\hat{\Lambda}$) and classical
fluctuations (here $\psi_\perp^{(1)}$) around the
Gross-Pitaevskii condensate field $\sqrt{N} \phi$, evolve according to the
same equations.
We find interestingly that the average
$\langle \psi_\perp^{(2)}\rangle$ in (\ref{eq:id_phi2})
evolves under the influence
of the mean field of the noncondensed particles, i.e. the Hartree-Fock
term and the anomalous average contribution. 
In the Wigner representation the Hartree-Fock mean field term
$2g \langle \psi_\perp^{(1)*} \psi_\perp^{(1)} \rangle$ differs from
the physical mean field
$2g \langle \hat{\Lambda}^\dagger \hat{\Lambda} \rangle$ by the term
$g(1-|\phi|^2 dV)/dV \simeq g/dV$. We note however that
this brings in a global phase change of the condensate wavefunction
having no effect on the one-body density matrix, and which is  
compensated anyway by the $-g/dV$ term in the Wigner drift term
(\ref{eq:drift}).
In our calculations
this is reflected by the fact that this term does not contribute
to $\phi_\perp^{(2)}$.

With the identities (\ref{eq:id_psi0}-\ref{eq:id_phi2})
we identify line by line the quantum expression (\ref{eq:rho1quant}) and the
truncated Wigner expression (\ref{eq:rho1wig}) for the one-body density
matrix of the system up to terms of $O(1)$:
these two expressions coincide apart from the term $1/2$ in
the occupation number of the mode $\phi$. This slight difference
$(1/2 \ll N)$ comes from the fact that in the initial sampling of the Wigner
function in thermal equilibrium we have treated classically the condensate
mode. 
These results establish the equivalence between the truncated Wigner
method and the time-dependent Bogoliubov approach of \cite{CastinDum}
up to neglected terms $O(1/\sqrt{N})$ in the limit (\ref{eq:limit}).

Let us however come back to the expansions performed in the limit
(\ref{eq:limit}). We have mentioned that the small formal parameter 
is $1/\sqrt{N}$ but we now wish to identify the small physical parameter of the expansion.
In the quantum theory of \cite{CastinDum} one gets the small parameter
\begin{equation}
\epsilon_{\rm quant} = \left(\frac{\langle \delta\hat{N}\rangle}{N}\right)^{1/2}
\end{equation}
where $\langle \delta\hat{N}\rangle$ is the Bogoliubov prediction 
for the number of noncondensed particles. 
In the expansion (\ref{eq:expansion}) of the evolving classical field
we compare the norm of the first two terms, ignoring the field phase change $\xi^{(1)}\phi$:
\begin{equation}
\epsilon_{\rm wig}=\left(\frac{\langle dV \sum_r
        |\psi^{(1)}_\perp|^2 \rangle}{N}\right)^{1/2}=
                   \left(\frac{\langle \delta \hat{N} \rangle +
                   ({\cal N}-1)/2}{N}\right)^{1/2}.
\end{equation}
The validity condition of the expansion (\ref{eq:expansion}) in the truncated Wigner
approach is then:
\begin{equation}
N \gg \langle \delta \hat{N} \rangle \;,\; { \cal N}/2 
\label{eq:necess1}
\end{equation}
which is more restrictive than in the quantum case.
What indeed happens in the regime $\langle \delta \hat{N} \rangle \ll N
< {\cal N}/2$? We expect the truncated Wigner approach not to recover the predictions
of the Bogoliubov approach of \cite{CastinDum} which are correct in this
limit. We therefore set a necessary condition for the validity
of the truncated Wigner approach:
\begin{equation}
N \gg  { \cal N}/2 .
\label{eq:necess2}
\end{equation}
We interpret this condition as follows: the extra noise introduced
in the Wigner representation (see discussion
after (\ref{eq:distriG})) contributes to the nonlinear term $g |\psi|^2$
in the evolution equation for the field; (\ref{eq:necess2})  means that
this fluctuating additional mean field potential of order
$g/(2 dV)$ should be much smaller than the condensate mean field of order
$g N/V$ where $V={\cal N} dV$ is the volume of the system.
Condition (\ref{eq:necess2}) is also equivalent to $\rho dV\gg 1$, where $\rho$
is the atomic density,
i.e. there should be on average
more than one particle per grid site.
We note that it is compatible with the conditions
(\ref{eq:dxassezpetit}) on the spatial steps of the grid in the regime of a degenerate ($\rho \lambda^3\gg 1$) and a weakly interacting
($\rho \xi^3 \gg 1$) Bose gas. Condition (\ref{eq:necess2}) is therefore
generically not restrictive.

A last important point for this subsection is that the time-dependent
Bogoliubov
approach, relying on a linearisation of the field equations around a pure
condensate solution, is usually restricted to short times in the case of an excited condensate,
so it cannot be used to test the condition of validity of the truncated
Wigner approach
in the long time limit.
It was found indeed in \cite{depletion} that 
nonlinearity effects in the condensate motion can lead to a polynomial or even
exponential increase in time of $\langle \delta\hat{N}\rangle$ which eventually invalidates
the time-dependent Bogoliubov approach. The truncated Wigner approach in
its full nonlinear
version does not have this limitation however, as we have checked with a
time-dependent
1D model in \cite{PRL}.

\subsection{Beliaev-Landau damping in the truncated Wigner approach}
\label{sub:Landau}

In this section we consider a spatially homogeneous Bose condensed gas 
in a cubic box in three dimensions with periodic boundary conditions. 
We imagine that with a Bragg scattering technique we excite 
coherently a Bogoliubov mode of the stationary Bose
gas, as was done experimentally at MIT \cite{Bragg_MIT1,Bragg_MIT2}, 
and we study how the excitation decays
in the Wigner approach due to Landau and Beliaev damping.  

\subsubsection{Excitation procedure and numerical results}
\label{sub:excitation}
We wish to excite coherently the Bogoliubov mode of
wavevector $k_0\neq0$.
With a Bragg scattering technique using two laser beams 
with wave vector difference $q$ and frequency difference $\omega$ 
we induce a perturbation potential
\begin{equation}
W=\int \, d^3r \, \left( \frac{W_0}{2}\,e^{i(q\cdot r -\omega t)} 
	+ c.c. \right)
\label{eq:perturb}
\end{equation}
We match the 
wavevector and frequency of the perturbation to the wavevector $k_0$
and the eigenfrequency $\omega_0=\epsilon_0/\hbar$ of the
Bogoliubov mode we wish to excite:
\begin{equation}
q=k_0	\hspace{2cm} \omega=\epsilon_0 /\hbar = \omega_0.
\label{eq:choix_param}
\end{equation}
During the excitation phase, we expect that two Bogoliubov modes 
are excited from the condensate, the modes with wavevectors
$k_0$ and $-k_0$. We anticipate the perturbative approach of next subsection
which predicts
that the mode of wavevector $k_0$,
being excited resonantly, has an amplitude growing
linearly with time, while
the mode with wavevector $-k_{0}$, being excited off-resonance,
has an oscillating amplitude
vanishing periodically when $t$ is a multiple integer of $\pi/\omega_0$.
In the truncated Wigner simulation we therefore stop the excitation
phase at 
\begin{equation}
t_{\rm exc}=\frac{\pi}{\omega_0} \,.
\end{equation}

We introduce the amplitudes of the classical field $\psi$
of the Bogoliubov modes. We first
define the field
\begin{equation}
\Lambda_{\rm static}(r) \equiv \frac{1}{\sqrt{N}} a_{\phi}^* \psi_\perp (r)
\label{eq:classLambda}
\end{equation}
where $a_\phi$ and $\psi_\perp$ are the components of $\psi$
orthogonal and parallel to the static condensate wavefunction
$\phi(r)=1/L^{3/2}$ (see (\ref{eq:splite})).
The component along the Bogoliubov mode with wavevector $k$ is then
\begin{equation}
b_k = dV \sum_r u_k^*(r) \Lambda_{\rm static}(r) - 
	v_k^*(r) \Lambda_{\rm static}^*(r) \,.
\end{equation}
The functions $u_{k}$ and $v_{k}$  are plane waves with wavevector
$k\neq0$ 
\begin{equation}
u_k(r)=\frac{1}{\sqrt{L^3}} U_k e^{i k\cdot r}  \hspace{3cm}
v_k(r)=\frac{1}{\sqrt{L^3}} V_k e^{i k\cdot r}
\label{eq:staticBogolmodes}
\end{equation}
and the real coefficients $U_k$ and $V_k$ are normalised to
$U_k^2-V_k^2=1$:
\begin{equation}
U_k + V_k =\frac{1}{U_k-V_k} = 
\left(\frac{\hbar^2k^2/2m}{\hbar^2k^2/2m+2\mu}\right)^{1/4}
\end{equation}
where the chemical potential is $\mu=gN/L^3$.

\begin{figure}[htb]   
\centerline{\epsfxsize=16cm \epsfbox{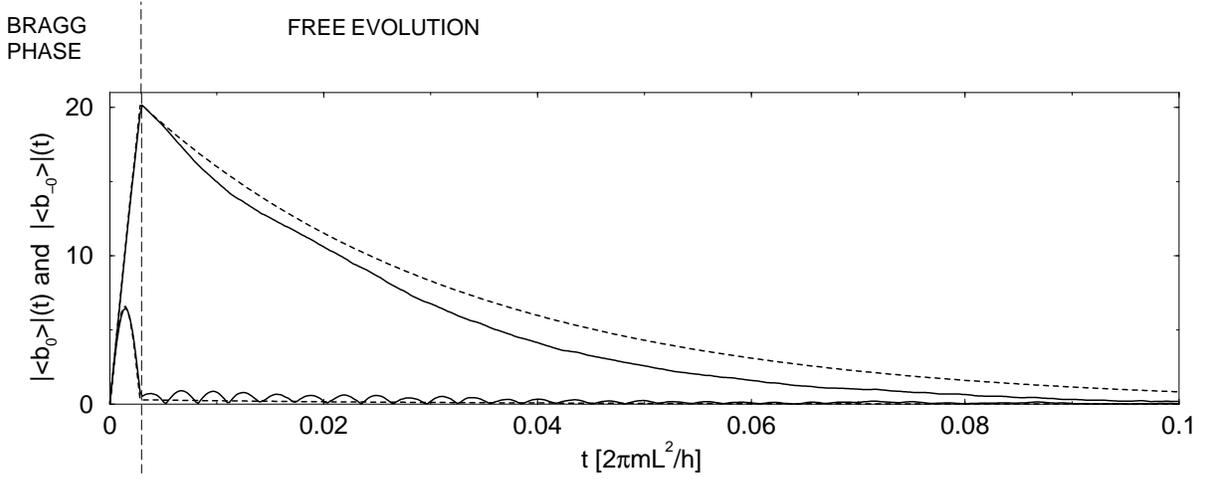}}
\caption{
\label{fig:landau}
Bragg excitation of a Bogoliubov mode of wavevector $k_0$ and frequency
$\omega_0$ for a finite temperature Bose condensed gas in a cubic
box.  The vertical dashed line at time $t=\pi/\omega_0$ indicates the time
after which the perturbation $W$ is discontinued.
Solid lines: evolution of the field amplitudes of the Bogoliubov
modes with wavevectors $k_0=(12\pi/L,0,0)$ (upper curve) and $-k_0$ 
(lower curve) in the Wigner simulation after averaging over 100
realizations.
Only the mode $k_0$ is excited resonantly by Bragg scattering.
After the coherent excitation Bragg phase, the amplitudes of the two modes
are damped. Dashed line: perturbative approach of 
subsection \ref{subsubsec:perturb}.  The truncated Wigner approach and
the perturbation theory give comparable results.
$N=5\times 10^4$, $k_B T=3 \mu$, $\hbar\omega_0=2.2 \mu$, $W_0=0.175\mu$, 
$\mu=500 \hbar^2 / m L^2$. 
In the Wigner simulation a grid 
with 22 points per dimension is used, so that ${\cal N}=22^3=10648\ll N$.
In the perturbative approach a grid of 48 points per dimension is used to
avoid truncation effects. 
The initial mean number of noncondensed
particles is $N-\langle N_0 \rangle\simeq 5000$.}
\end{figure}

We denote by $b_0$ the amplitude of the field $\Lambda_{\mbox{\scriptsize
static}}$
along the Bogoliubov mode of wavevector $k_0$, and $b_{-0}$ the amplitude
along the mode with opposite wavevector.
We show the mean values of these amplitudes
as function of time obtained from the truncated Wigner simulation 
in figure \ref{fig:landau}. In the initial thermal state these mean
values vanish, and they become nonzero during the excitation
phase due to the coherent excitation procedure. At later times
they decay to zero again \cite{singlereal}.

\subsubsection{Perturbative analysis of the truncated Wigner approach:
Beliaev-Landau damping }
\label{subsubsec:perturb}
In the appendix \ref{app:eqlambda_ex} we report the
exact equations of motion
of the classical field $\Lambda_{\rm static}$ defined by (\ref{eq:classLambda})
in the truncated Wigner approach. 
We now make the assumption that $\Lambda_{\rm static}$ is small 
compared with $\sqrt{N}\phi$, implying that
\begin{equation}
N \gg \langle \delta \hat{N} \rangle \; , \;  \frac{{\cal N}}{2}
\end{equation}
where $\langle \delta \hat{N} \rangle$ represents here the mean number of
particles  in the excited modes of the cubic box. In this regime   
we neglect terms which are at least cubic in $\Lambda_{\rm static}$
in (\ref{eq:lambdaTW}) and we replace
the number of particles in the ground state of the box by the total number
of particles $N$, except in the zeroth order term in $\Lambda_{\rm static}$ 
where we replace it by its initial mean value $\langle N_0 \rangle$.
We then find:
\begin{eqnarray}
i\hbar\frac{d}{dt} \Lambda_{\rm static}\simeq &&\sqrt{\langle N_0 \rangle} 
{\cal Q} h_0 \phi + 
{\cal Q} h_0 \Lambda_{\rm static} +
\frac{Ng}{L^3}(\Lambda_{\rm static}^*+2\Lambda_{\rm static}) \nonumber \\
&+& \frac{g\sqrt{N}}{\sqrt{L^3}}{\cal Q}(\Lambda_{\rm static} \Lambda_{\rm static} 
+2 \Lambda_{\rm static}^* \Lambda_{\rm static}) -\frac{1}{\sqrt{N L^3}}
\Lambda_{\rm static}(r)dV\sum_s W_0 \cos(q\cdot s-\omega t)
\Lambda_{\rm static}^*(s)\,,
\label{eq:simpleLambda}
\end{eqnarray}
where $W_0$ is non zero only during the excitation phase.
In this equation $h_0=p^2/2m + W_0 \cos(q \cdot r-\omega t)$ is the one-body part 
of the Hamiltonian including the
kinetic energy and the Bragg excitation potential, and
${\cal Q}$ projects orthogonally to the static condensate mode $\phi$.
The term of zeroth order in $\Lambda_{\rm static}$ is a source term which causes 
$\Lambda_{\rm static}$ to acquire a nonzero mean value during the evolution.
The terms of first order in $\Lambda_{\rm static}$ in (\ref{eq:simpleLambda})
describe the evolution in the static Bogoliubov approximation. 
Terms of second order provide the damping we are looking for. 
We project equation (\ref{eq:simpleLambda}) 
over the static Bogoliubov modes (\ref{eq:staticBogolmodes}) by using:
\begin{equation}
\Lambda_{\rm static}(r)= \sum_{k\neq0} b_{k} u_{k}(r) + b_k^* v_{k}^*(r)
\end{equation}
with the mode functions
$u_{k}(r)$ and $v_{k}(r)$ defined in (\ref{eq:staticBogolmodes}).
Terms nonlinear in $\Lambda_{\rm static}$ in (\ref{eq:simpleLambda}) then
correspond to an interaction between the Bogoliubov modes.

We assume that the excitation phase is much shorter than the damping time
of the coherently excited mode. As a consequence 
we can neglect in this phase the processes involving interaction among the 
Bogoliubov modes. 
Also in the action of the perturbation $W$ we keep only the term acting on
the condensate mode, that is the first term on the right hand side of (\ref{eq:simpleLambda}),
which is $\sqrt{\langle N_0\rangle}$ larger than the 
terms acting on the noncondensed modes.
For the choice of parameters (\ref{eq:choix_param})
only the two modes with wavevectors
$k_0$ and $-k_0$ are excited from the condensate
by the perturbation $W$; the amplitudes of the field in these modes
evolve according to
\begin{eqnarray}
\label{eq:lin1}
i \hbar \frac{d}{dt} b_0 &=& \hbar \omega_0 b_0 + \sqrt{\langle N_0\rangle} \frac{W_0}{2}
        (U_0+V_0) \, e^{-i \omega_0 t} \\
i \hbar \frac{d}{dt} b_{-0} &=& \hbar \omega_0 b_{-0} +
                \sqrt{\langle N_0\rangle} \frac{W_0}{2}
        (U_0+V_0) \, e^{i \omega_0 t}\,.
\label{eq:lin2}
\end{eqnarray}
By integrating these equations we realise that the mean amplitude 
$\langle b_0 \rangle$ grows linearly in time, since the mode is excited
resonantly,
while the mean amplitude $\langle b_{-0}\rangle$ oscillates and vanishes at
$t=\pi/\omega_0$.

After the excitation phase we include the second order terms that provide
damping:  
\begin{eqnarray}
i \hbar \frac{d}{dt} {b}_{0} &=& \epsilon_0 b_0 + 
\sum_{i,j}  A_{i,j}^0 {b}_i {b}_j +
            (A_{i,0}^j+A_{0,i}^j) b_i^* b_j 
+ \sum_{i,j} (B_{i,j,0}+B_{0,i,j}+B_{i,0,j}) b_i^* b_j^* 
\label{eq:b0}
\end{eqnarray}
with
\begin{eqnarray}
A_{j,k}^i&=&\frac{g \sqrt{N}}{L^3}
	[U_i(U_j+V_j)U_k+(U_i+V_i)V_j U_k+V_j(U_k+V_k)V_i]\delta_{i,j+k} \\
B_{i,j,k}&=&\frac{g \sqrt{N}}{L^3}
	V_i(U_j+V_j)U_k \delta_{-i,j+k}  \,.
\end{eqnarray}
and where $i,j,k$ denote momenta.
The last terms with the $B$'s in (\ref{eq:b0}) do not conserve the Bogoliubov
energy and we can neglect them here for the calculation of the damping rate since 
we are going to use second order perturbation theory; we would have to keep them
in order to calculate frequency shifts.
In the terms with the $A$'s we recognise two contributions: 
the term with $A_{i,j}^0$ describes a Beliaev process where the excited
mode can decay into two different modes while the term with 
$A_{i,0}^j+A_{0,i}^j$ describes a Landau process where the excited
mode by interacting with another mode is scattered into a third mode 
\cite{quantique_aussi}.
We introduce the coefficients $\tilde{b}$ in the interaction picture 
\begin{equation}
\tilde{b}_j=b_j \, e^{i \epsilon_j \,t/\hbar}
\end{equation}
where $\epsilon_j$ is the Bogoliubov eigenenergy of the mode
with wavevector $j$,
and we solve (\ref{eq:b0}) to second order of time-dependent
perturbation theory to obtain:
\begin{eqnarray}
\langle \tilde{b_0}(t) - \tilde{b_0}(0) \rangle & \simeq &
- \frac{1}{\hbar^2} 
  \sum_{i,j} A_{i,j}^0 (A_{i,j}^0+A_{j,i}^0) \,
	I_t(\epsilon_0-\epsilon_i-\epsilon_j)
        (1+\bar{n}_i+\bar{n}_j) \langle \tilde{b_0}(0) \rangle \nonumber \\
& &  - \frac{1}{\hbar^2}
  \sum_{i,j} (A_{i,0}^j+A_{0,i}^j)^2 \,
	I_t(\epsilon_0+\epsilon_i-\epsilon_j)
        (\bar{n}_i-\bar{n}_j) \langle \tilde{b_0}(0) \rangle \nonumber \\
& & - \frac{1}{\hbar^2} 
    2 (A_{0,0}^{0+0})^2  \,
	I_t(\epsilon_0+\epsilon_0-\epsilon_{0+0})
	\langle \tilde{b_0^*}(0) \tilde{b_0}(0) \tilde{b_0}(0) \rangle
\label{eq:lunga}
\end{eqnarray}
where $0+0$ represents the mode of wavevector $2k_0$ and where
\begin{eqnarray}
I_t(\nu)&=&\int_0^t d\tau \, e^{i \nu \tau /\hbar} \, f_\tau(\nu) \\
f_\tau(\nu)&=& \int_0^\tau d\theta \, e^{-i \nu \theta /\hbar} .
\end{eqnarray}
The $\bar{n}_j$'s are the occupation numbers of the Bogoliubov modes
in thermal equilibrium given by the Bose formula 
\begin{equation}
\bar{n}_j=\frac{1}{e^{\epsilon_j/k_B T}-1}
\end{equation}
where $\epsilon_j$ is the energy of the Bogoliubov mode.
In the language of nonlinear optics
the last line in (\ref{eq:lunga})
describes a $\chi_2$ effect or a second harmonic generation which can
be important if the conservation of energy condition $\epsilon_{2 k_0}=2
\epsilon_{k_0}$ is satisfied and if the initial amplitude 
$\langle \tilde{b_0}(0) \rangle = \beta$ is large since one has
\begin{equation}
\langle \tilde{b_0^*}(0) \tilde{b_0}(0) \tilde{b_0}(0) \rangle =
|\beta|^2 \beta + \bar{n}_0 2 \beta\,.
\end{equation}
We have checked that the $\chi_2$ effect is negligible for the low amplitude
coherent excitations considered in the
numerical examples of this paper: $\epsilon_0$ is larger than $\mu$ so that
$k_0$ is not in the linear part of the Bogoliubov spectrum and therefore the second
harmonic generation process is not resonant. 
By using the fact that:
\begin{equation}
\mbox{Re} \; I_t(\nu) = \frac{1}{2} |f_t(\nu)|^2 
=\frac{2\hbar^2}{\nu^2}\sin^2\frac{\nu\tau}{2\hbar}
\equiv \pi \hbar t \delta_t(\nu)
\end{equation}
where $\delta_t(\nu)$ converges to a Dirac delta distribution
in the large $t$ limit,
we calculate the evolution of the modulus of the Bogoliubov mode amplitude
\begin{eqnarray}  
\frac{|\langle b_0(t)\rangle|-|\langle {b_0}(0)\rangle|} 
{|\langle {b_0}(0)\rangle|}
&\simeq&
- \frac{\pi t}{\hbar} \,
	\sum_{i,j} A_{i,j}^0(A_{i,j}^0+A_{j,i}^0) \,
	\delta_t(\epsilon_0-\epsilon_i-\epsilon_j)
         (1+\bar{n}_i+\bar{n}_j) \nonumber \\
	 && - \frac{\pi t}{\hbar} \,
	\sum_{i,j} (A_{i,0}^j+A_{0,i}^j)^2 \,
	\delta_t(\epsilon_0+\epsilon_i-\epsilon_j)
         (\bar{n}_i-\bar{n}_j)  \,.
\label{eq:gamma}
\end{eqnarray}
This formula can be applied to a finite size box as it contains
finite width $\delta$'s. By plotting equation (\ref{eq:gamma}) as a function 
of time we can identify a time interval over which it
is approximately linear in time, and we determine
the slope
$-\gamma_{\mbox{\scriptsize perturb}}$ with a linear fit
\cite{intervalle}.
Heuristically we then compare $\exp(-\gamma_{\mbox{\scriptsize perturb}} t)$ 
to the result of the truncated Wigner simulation, 
see figure \ref{fig:landau} and we obtain a good agreement for this particular
example \cite{gamma_aussi}.

In the thermodynamic limit, when the Bogoliubov spectrum becomes
continuous, 
the discrete sums in (\ref{eq:gamma}) can be replaced by integrals
and the finite width $\delta_t$ is replaced by a Dirac $\delta$ 
distribution. In this case an analytical expression for the damping rate 
can be worked out and we recover {\sl exactly} the expression for
the Beliaev and Landau damping rate obtained in the quantum
field theory \cite{Vincent,Stringari_Pitaevskii,Shlyapnikov}.

\subsubsection{Validity condition of the truncated Wigner approach}

\begin{figure}[htb]
\centerline{\epsfxsize=7cm \epsfbox{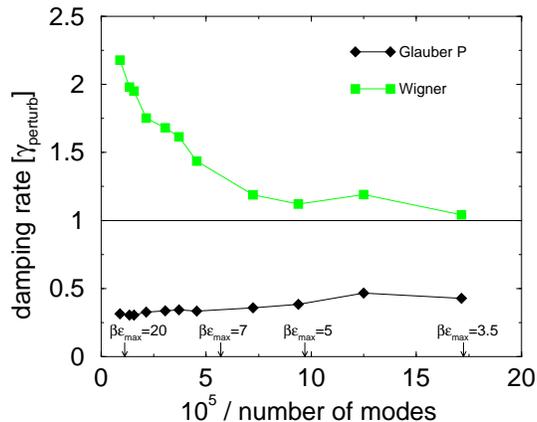}}

\caption{Damping rate of the coherent excitation in the Bogoliubov mode
of wavevector $k_0=(12\pi/L,0,0)$ and of frequency $\omega_0$
as a function of the inverse number of modes in the
grid $1/{\cal N}$ for the Glauber-P and the Wigner
distributions.
Each disk represents the average over 100 realisations of the simulation
and the lines are a guide to the eye. $N=10^5$, $k_B T=3 \mu$,
$\mu=500\hbar^2/mL^2$, so that $\hbar\omega_0=2.2\mu$,
$\gamma_{\mbox{\scriptsize perturb}}^{-1}=0.061 m L^2/\hbar$, $W_0=0.0874
\mu$.
The damping rate is expressed in units of $\gamma_{\mbox{\scriptsize
perturb}}$. Arrows indicate some values of $\epsilon_{\rm max}/k_B T$
where $\epsilon_{\rm max}$ is the maximal Bogoliubov energy on the
grid.} \label{fig:effet_coupure} 
\end{figure}

We now investigate numerically the influence of the grid size on
the predictions of the truncated Wigner simulation.
The line with squares in figure \ref{fig:effet_coupure} shows the damping 
rate obtained from the Wigner simulation,
defined as the inverse of the $1/e$ half-width of $|\langle b_0(t)\rangle|$,
as a function of the inverse grid size $1/{\cal N}$.
For small grids
the results of the simulations reach a plateau close to the perturbative
prediction $\gamma_{\rm perturb}$. 
For large grids the damping rate in the simulation
becomes significantly larger than $\gamma_{\rm perturb}$.
Since the perturbative prediction reproduces the known result for
Beliaev-Landau damping, we conclude that the results of the truncated
Wigner simulation become incorrect for large grid sizes.
The reason of such a {\it spurious} damping appearing in the Wigner
simulation for large ${\cal N}$ will become clear below.

It is tempting to conclude from the perturbative calculation of subsection
\ref{subsubsec:perturb} that the validity condition of the truncated
Wigner approach is dictated only by the condition ${\cal N}\ll N$. To check
this statement we have performed a second set of simulations (not shown)
for a particle number $N$ reduced by a factor of two keeping the size of
the box $L$, the chemical potential $\mu=Ng/L^3$ and the temperature
fixed.  If the condition of validity of the truncated Wigner approach
involves only the ratio $N/{\cal N}$ the plateaux in the damping time
should start at the same value of $N/{\cal N}$ for the two sets of
simulations.  However this is not the case, and we have checked that on
the contrary, the two curves seem to depend on the number of modes only.

\begin{figure}[htb]
\centerline{\epsfxsize=7cm \epsfbox{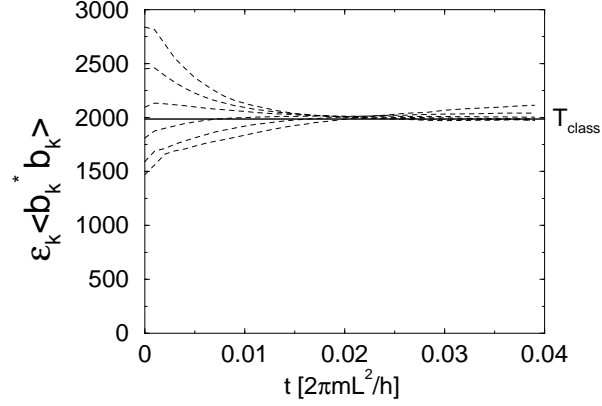}}
\caption{\label{fig:thermalisation}
Evolution of the squared amplitudes $\langle b_k^* b_k\rangle$ 
of the classical field Bogoliubov modes multiplied by the 
corresponding Bogoliubov energy $\epsilon_k$
in the truncated Wigner simulation
in the absence of the Bragg perturbation.
We have collected the Bogoliubov modes in energy channels of width
$2\mu$, so that the plotted quantity is the average among each channel
of $\epsilon_k\langle b_k^* b_k\rangle$, with increasing energy from top
to bottom at initial time $t=0$. 
The thick horizontal line is the expected temperature
$T_{\rm class}$ of the equilibrium classical field distribution 
as given by (\ref{eq:Tclass}). Parameters are:
$N=5\cdot 10^4$, $k_B T=3\mu$, $\mu=500\hbar^2/mL^2$ and the
vertical axis of the figure is in units of $\hbar^2/mL^2$, where $L$ is the cubic
box size. The number of modes is 22 per spatial dimension so that
the maximum Bogoliubov energy allowed on the grid is $\epsilon_{\rm max}=15.3 \mu$. 
The averaging in the simulation is performed over 500 realisations.
}
\end{figure}

Another way to put it is that the condition to have agreement between the
truncated Wigner simulation and the perturbation theory of section
\ref{subsubsec:perturb} is not (or not only) that the number of particles
should be larger than the number of modes. There is in fact another ``hidden''
condition in the perturbative calculation which is the hypothesis that the
occupation numbers of the Bogoliubov modes are constant during the evolution.  
In reality, even in absence of the Bragg perturbation, our initial state which
reproduces the correct thermal distribution for the quantum Bose gas, is not
stationary for the classical field evolution (\ref{eq:GPE}).  The perturbative
expression (\ref{eq:gamma}) holds indeed in the limit $N/{\cal N}\gg 1$, but the
occupation numbers of the Bogoliubov modes, initially equal to the Bose formula
$\bar{n}_j$, change in the course of the time evolution in the simulation and
this affects the damping rate.  This effect is neglected in the perturbative
formula (\ref{eq:gamma}) and it is found numerically to take place on a time
interval comparable to the damping time of the Bogoliubov coherent excitation as
we show in figure \ref{fig:thermalisation}.

What it is expected to happen in the absence of external perturbation
is that the classical field equation 
(\ref{eq:GPE}), in the three-dimensional cubic box geometry considered here,
displays an ergodic behaviour leading to thermalisation of the classical
field $\psi$ towards its equilibrium distribution \cite{Davis,Polonais}.
In the regime where the noncondensed fraction is small and the number of modes
is smaller than $N$, we can approximately view the classical field
as a sum of Bogoliubov oscillators $b_k$ weakly coupled
by terms leading to the nonlinearities
in (\ref{eq:lambdaTW}).
In the equilibrium state for the classical field dynamics we then expect
the
occupation numbers of the Bogoliubov modes to be given by the equipartition
formula:
\begin{equation}
\langle b_k^* b_k\rangle_{\rm class} = 
\frac{k_B T_{\rm class}}{\epsilon_k} \label{eq:classicaldistribution}
\end{equation}
attributing a mean energy of $k_B T_{\rm class}$ to each of the Bogoliubov mode.
The classical field equilibrium temperature $T_{\rm class}$ can then be deduced
from the approximate conservation of the Bogoliubov energy \cite{check_ener}:
\begin{eqnarray}
k_B T_{\rm class} &=& \frac{1}{{\cal N}-1} \sum_{k\neq 0} \epsilon_k
\langle b_k^* b_k\rangle(t=0) \nonumber \\
&=& \frac{1}{{\cal N}-1} \sum_{k\neq 0} \left[\frac{\epsilon_k}{\exp(\beta\epsilon_k)
-1} +\frac{1}{2}\epsilon_k\right] \label{eq:Tclass_demi}\\
&=& \frac{1}{{\cal N}-1} \sum_{k\neq 0} \frac{\epsilon_k}{2\tanh(\beta\epsilon_k/2)}\,.
\label{eq:Tclass}
\end{eqnarray}
The thermalisation of the Bogoliubov modes to the new temperature 
$T_{\rm class}$ 
is nicely demonstrated in figure \ref{fig:thermalisation}. 
One sees that $\epsilon_k \langle b_k^* b_k\rangle$ indeed converges
to a constant value almost independent of $k$.
From the fact that $\tanh x < x$ for any $x>0$ we deduce that the classical
equilibrium temperature $T_{\rm class}$ is always larger than the real
physical temperature $T$ of the gas. In the regime
$k_B T \gg \mu$ this \lq heating' increases 
the squared amplitudes $\langle b_k^* b_k\rangle$ of the modes of energy
$\sim \mu$ by a factor $\simeq
T_{\rm class}/T$. Since the Landau damping rate is approximately
proportional
to the populations of these modes
\cite{Vincent,Stringari_Pitaevskii,Shlyapnikov}
the damping rate is increased roughly by a factor $T_{\rm class}/T$,
an artifact of the truncated Wigner approximation.

It is clear that $T_{\rm class}$ will remain very close to $T$ as long as the 
maximum Bogoliubov energy allowed in the simulation remains 
smaller than $k_BT$.
One can indeed in this case expand (\ref{eq:Tclass}) 
in powers of $\beta\epsilon_k$. 
One has to expand the hyperbolic tangent up to cubic
order to get a nonzero correction:
\begin{equation}
\frac{T_{\rm class}}{T} \simeq
1+\frac{1}{{\cal N}-1}\sum_{k\neq 0} \frac{(\beta\epsilon_k)^2}{12}.
\label{eq:Tclasslow}
\end{equation}
The absence of terms of order $\beta\epsilon_k$ in
(\ref{eq:Tclasslow}) is a fortunate consequence of the noise
added to the field in the Wigner representation. This added noise shifts the average 
$\langle b_k^* b_k\rangle(t=0)$ by $1/2$ with respect to the Bose formula.

\begin{figure}[htb]
\centerline{\epsfxsize=7cm \epsfbox{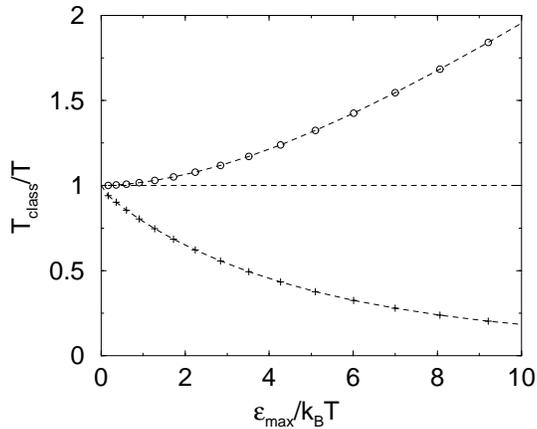}}
\caption{
Equilibrium temperature $T_{\rm class}$
of the classical gas as function of the maximum energy $\epsilon_{\rm max}$
of the Bogoliubov
modes on the momentum grid with the assumption of equipartition of the
energy in the
Bogoliubov modes. Circles: the initial field distribution
is the Wigner distribution for the quantum gas at temperature $T$.
Crosses: Glauber-P distribution defined in \protect\cite{Modugno}, amounting to
the removal of the added Wigner noise from the initial field
distribution. The dashed lines are a guide to the eye. The number
of momentum components along each dimension of space goes from 2
to 30 in steps of 2. The chemical potential is $\mu=500\hbar^2/mL^2$
and the temperature is $k_B T=3\mu$.
\label{fig:Tclass}}
\end{figure}

When the maximum Bogoliubov energy becomes much larger than $k_B T$ we
expect $T_{\rm class}$ to become significantly larger than $T$. This is
illustrated in figure \ref{fig:Tclass} obtained by a numerical calculation
of the sum in (\ref{eq:Tclass}) for increasing grid sizes. We have also
plotted in this figure the value that one would obtain for $T_{\rm class}$
in the absence of the added Wigner noise (i.e. in a Glauber-P approach),
that is by removing the terms $\epsilon_k/2$ in (\ref{eq:Tclass_demi}).
The Glauber-P distribution for the field $\psi$ in the sense of \cite{Modugno} is given by
\begin{equation}
\psi=N_0 \phi+ \sum_{k\neq 0} b_k u_k+b_k^* v_k^* \label{eq:psiglauber}
\end{equation}
where the $b_k$ are chosen from a Gaussian distribution such that $\langle
b_k^* b_k \rangle=1/({\rm exp}(\beta \epsilon_k)-1)$ and the value of
$N_0$ is dictated by the normalisation condition $||\psi||^2=N$. In this
case $T_{\rm class}$ is always smaller than $T$, and deviates from $T$ for
smaller grid sizes, since the fortunate cancellation of the order
$\beta\epsilon_k$ obtained in (\ref{eq:Tclasslow}) does not occur anymore.
We expect in this case a spurious reduction of the damping rate. We have
checked it by evolving an ensemble of fields of the form
(\ref{eq:psiglauber}) with the Gross-Pitaevskii equation and we found that
the damping rate is always smaller than half of the correct result even
for the smallest grids that we tested,  see the line with diamonds in figure
\ref{fig:effet_coupure}. 

\section{Conclusion}
We have considered a possible way of implementing
the truncated Wigner approximation
to study the time evolution of trapped
Bose-Einstein condensates perturbed from
an initial finite temperature equilibrium state.
First a set of random classical fields $\psi$ is generated to approximately
sample the initial quantum thermal equilibrium state of the gas, in the Bogoliubov
approximation assuming a weakly interacting and almost pure
Bose-Einstein condensate.
Then each field $\psi$ is evolved in the classical field
approximation, that is according to the
time-dependent Gross-Pitaevskii equation, with the crucial difference
with respect to the more traditional use of the Gross-Pitaevskii
equation that the field $\psi$ is now the whole matter field rather
than the field in the mode of the condensate.

The central part of this paper is the investigation of the validity
conditions of this formulation of the truncated Wigner approximation.

For short evolution times of the fields $\psi$ the dynamics of the
noncondensed modes, i.e.\ the
components of the field orthogonal to the condensate mode,
is approximately
linear; we can then use the time-dependent Bogoliubov approximation, both
for the exact quantum problem and for the truncated Wigner approach. 
A necessary condition for  the
truncated Wigner approach to correctly reproduce the quantum results
is then
\begin{equation}
N \gg {\cal N}/2 
\end{equation}
where ${\cal N}$ is the number of modes in the Wigner approach and $N$
is the total number of particles in the gas. This condition can in
general be satisfied in the degenerate and weakly interacting regime
without introducing truncation effects due to a too small number of modes.

For longer evolution times the nonlinear dynamics of the noncondensed
modes comes into play. When the classical field
dynamics generated by the Gross-Pitaevskii equation is ergodic,
e.g.\ in the example of a three dimensional gas in a cubic box considered
in this paper, the set of Wigner fields $\psi$ evolves from the
initial distribution mimicking the thermal state of the quantum gas
at temperature $T$ to a classical field equilibrium distribution
at temperature $T_{\rm class}$.
Since noise is added in the Wigner representation in all modes
of the classical field to mimic quantum fluctuations
it turns out that $T_{\rm class}$ is always larger
than $T$. If $T_{\rm class}$ deviates too much from $T$ the truncated Wigner
approximation can give incorrect predictions. For example we have found
that the Beliaev-Landau damping of a Bogoliubov mode in the box,
taking place with a time scale comparable to that of the \lq thermalisation'
of the classical field, is accelerated in a spurious way as the classical
field \lq warms up'. A validity condition for the truncated Wigner
approach in this long time regime is therefore
\begin{equation}
|T_{\rm class} - T| \ll T. \label{eq:condition}
\end{equation}
This condition sets a constraint on the maximum energy of the Bogoliubov
modes $\epsilon_{\rm max}$ in the Wigner simulation: $\epsilon_{\rm max}$
{\it should not exceed a few} $k_B T$. More precisely one can use
the following inequality to estimate the error \cite{pourquoi}:
\begin{equation}
\frac{|T_{\rm class} - T |}{T} 
<\frac{1}{12} \frac{\langle\epsilon_k^2\rangle}{(k_B T)^2}
<\frac{1}{12} \left(\frac{\epsilon_{\rm max}}{k_B T}\right)^2
\end{equation}
where $\langle\epsilon_k^2\rangle$ is the arithmetic mean of the squares of all
the Bogoliubov energies in the Wigner simulation.

The fact that the initial set of Wigner fields is nonstationary under the
classical field evolution could be a problem: the time-dependence
of the observables could be affected in an unphysical way during the thermalisation
to a classical distribution of the ensemble.  To avoid this, we could
{\em start} directly from the thermal equilibrium classical distribution \cite{Davis,Burnett}, 
restricting to the regime $\epsilon_{\rm max} < k_B T$.

A remarkable feature of the Wigner simulation is that $T_{\rm class}$ deviates from $T$ at low values of
$\epsilon_{\rm max}$ only {\sl quadratically} in $\epsilon_{\rm max}/k_B T$.
This very fortunate feature originates from the added noise in the Wigner
representation. It explains why for $\epsilon_{\rm max}$ as high as 3.5 $k_B T$
the truncated Wigner approach can still give very good results for the
Beliaev-Landau damping time (see Fig. \ref{fig:effet_coupure}). In contrast,
if we remove the Wigner added noise, in the so-called Glauber-P representation,
or if we add more noise, in the so-called Q representation, $T_{\rm class}$
deviates from $T$ {\sl linearly} in $\epsilon_{\rm max}/k_B T$. In this case we
expect that the condition of validity of the classical Gross-Pitaevskii equation
will be that all modes in the problem must be highly occupied, resulting in the
stringent condition $\epsilon_{\rm max} < k_B T$. We
therefore conclude that the Wigner representation is the most favorable
representation of the quantum density operator with which to perform the
classical field approximation. This fact, known in
quantum optics for few mode systems, was not obvious for the highly multimode
systems that are the finite temperature Bose gases.

Still, condition (\ref{eq:condition}) is a serious limitation of the
truncated Wigner method for simulating general ergodic three dimensional
systems. One possibility to overcome this limitation
is to proceed as in \cite{Stoof,Fudge} i.e. to treat the high
energy modes as a reservoir, which leads to the inclusion of a stochastic
term in the Gross-Pitaevskii equation. The advantage of this treatment is
that the additional term has dissipative effects and thermalises the
system to the correct quantum field thermal distribution in the stationary
state as opposed to the classical one.
However, one of the conceptual advantages of the truncated Wigner method
and of classical field methods in general \cite{Kagan,Sachdev,Davis,Polonais}
which we would like to keep is that apparent damping and irreversibility
arise from the dynamics of a {\it conservative} equation (the
Gross-Pitaevskii or nonlinear Schr\"{o}dinger equation) as is the case in 
the original Hamiltonian equations for the quantum field.

Laboratoire Kastler Brossel is a research unit of \'Ecole Normale
Sup\'erieure and of Universit\'e Pierre et Marie Curie, associated to
CNRS. We acknowledge very useful discussions with Crispin Gardiner. This
work was partially supported by National Computational Science Alliance
under DMR 9900 16 N and used the NCSA SGI/CRAY Origin2000.

\appendix

\section{Bare vs effective coupling constant}
\label{appen:g0}
In this appendix we describe how to adjust the potential ${\cal V}(r)$
defined on the grid in the simulation in order to reproduce correctly 
the low energy scattering properties of the true interatomic potential.

We start with the Schr\"odinger equation for a scattering state
$\phi(r)$ of the discrete delta potential ${\cal V}(r)\equiv(g_0/dV)\delta_{r,0}$
on the spatial grid of size $L_\nu$ and volume $V$:
\begin{equation}
\epsilon\phi(r) = 
\left(\frac{p^2}{m}\phi \right) (r)+\frac{g_0}{dV}\phi(r)\delta_{r,0}
\end{equation}
where $m$ is twice the reduced mass and where $\phi(0)$
is different from zero.  We project this equation on plane
waves of momentum $k$:
\begin{equation}
\tilde{\phi}(k) = \frac{g_0}{V^{1/2}} \frac{\phi(0)}{\epsilon-\hbar^2k^2/m},
\end{equation}
where $\tilde{\phi}(k)$ is the component of $\phi$ on the plane wave
$e^{ik\cdot r}/\sqrt{V}$. Fourier transforming back gives $\phi(0)$; dividing
the resulting equation by $\phi(0)$ leads to the quantization condition
\begin{equation}
1=\frac{1}{V}\sum_k \frac{g_0}{\epsilon-\hbar^2k^2/m} \label{eq:Iacopo}.
\end{equation}

We define the effective coupling constant $g_{\rm eff}$ in such a way that
the energy of the lowest scattering state of the pseudopotential $g_{\rm
eff}\delta(r) \partial_r(r \, \cdot)$ in the box is the same as the energy
of the lowest scattering state solution of (\ref{eq:Iacopo}).

We now restrict ourselves to the case where the size of the box is much
larger than the scattering length associated with $g_{\rm eff}$. 
In this case the energy of the lowest
scattering state for the continuous theory with the
pseudopotential is very close to $g_{\rm eff}/V$, so that we can
calculate $g_{\rm eff}$ from the equation $\epsilon = g_{\rm eff}/V$.
In this large box case, one can then check that
the energy  $\epsilon$ is negligible as compared to 
$\hbar^2k^2/m$ except if $k=0$. This gives
\begin{equation}
g_{\rm eff} = \frac{g_0}{1+\frac{1}{V}\sum_{k\neq 0} \frac{g_0}{\hbar^2k^2/m}}
\end{equation}
which allows us to adjust $g_0$ in order to have $g_{\rm eff}=g\equiv 4\pi \hbar^2a/m$
where $a$ is the scattering length of the true interatomic potential.

The sum over $k$ in the denominator can be estimated by replacing the sum
by an integral over $k$ and is found to be on the order of $k_{\rm max} a_0$
where $g_0=4\pi\hbar^2a_0/m$ and $k_{\rm max}$ is the maximum momentum on the grid.
$g_0$ is therefore very close
to $g_{\rm eff}$ when condition (\ref{eq:noname}) is satisfied, so that we
can set $g_0\simeq g_{\rm eff}=g$.
In the opposite limit of a grid step size tending to zero one gets $g_{\rm eff}
\rightarrow 0$, and we recover the known fact that a delta potential does not
scatter in the continuous limit.
We would have to increase $g_0$ continuously up to infinity as the grid step size
tended to zero, if we wanted to get a finite $g_{\rm eff}$ in this limit.

\section{An improved Brownian motion simulation}
\label{app:stoch}

{\sl A better choice for $\alpha$ and $Y$} --
In our previous work \cite{Kuhtai} the drift matrix $\alpha$ and the
noise matrix $Y$ were the hyperbolic sine and cosine of ${\cal L}/(2 k_B
T)$,
which imposed a time step $dt$ in the simulation which was exponentially
small
in the parameter $\epsilon_{\rm max}/(k_B T)$, where $\epsilon_{\rm max}$
is the largest eigenvalue of $\cal L$ allowed on the spatial grid
of the simulation.
We have now identified a choice that does not have this disadvantage:
\begin{eqnarray}
\alpha &=& 2M \label{eq:mieux_alpha}\\
Y &=&
\left( \begin{tabular}{cc}
${\cal Q}$ & 0 \\
0 & ${\cal Q}^*$
\end{tabular} \right)
, \label{eq:mieux_Y}
\end{eqnarray}
where the projector ${\cal Q}$ is defined in (\ref{eq:defQ}).
With this new choice for $\alpha$ and $Y$
both the friction matrix and the noise matrix are bounded from
above by unity, which allows a much larger $dt$ in the case 
$\epsilon_{\rm max} > k_B T$. To calculate the action
of matrix $\alpha$ on the vector $(\psi_\perp,\psi_\perp^*)$ we write
the hyperbolic tangent as:
\begin{equation}
\tanh x = x \, \frac{\tanh x}{x} \equiv x F(x^2).
\end{equation}
The function $F(u)$ is then expanded on Chebyshev polynomials
in the interval $u\in [0, (\epsilon_{\rm max}/(2 k_B T))^2]$
and approximated by a polynomial of a given degree, typically
15 for $\epsilon_{\rm max}/(2 k_B T)=3$ 
and 25 for $\epsilon_{\rm max}/(2 k_B T)=6$, obtained by truncating a 
Chebyshev expansion of degree 50 \cite{NumRecCheb}.

\noindent{\sl An improved integration scheme} --
Initially we set $\psi_\perp= 0$. Since the noise $d\xi$ is Gaussian,
and because the stochastic differential equation (\ref{eq:brown}) is
linear, the probability distribution of $\psi_\perp$ is guaranteed to be
Gaussian at any step of the integration so that the issue of the convergence
of the distribution to the correct steady state distribution
(\ref{eq:Pperp}) can be discussed in terms of the convergence of
the covariance matrix of the distribution to its right steady state
value. Two issues in particular should be addressed:
the error introduced by the discretisation in time (finite time step
$dt$ of integration), and the error introduced by the integration over
a finite time interval (approach to the steady state distribution).

We now explain how to face the first problem with an efficient integration
scheme yielding an error on the steady state covariance matrix
of the distribution scaling as $dt^2$, rather than $dt$ for the
simple Euler scheme. In the numerical scheme 
the vector $\vec{X}\equiv (\psi_\perp,\psi_\perp^*)$ that stores the
values 
of the field $\psi_\perp$ and of its complex conjugate 
$\psi_\perp^*$ on the discrete grid obeys
the recursion relation:
\begin{equation}
\vec{X}_{[t=(n+1)dt]}= (1-\alpha_{\rm num}dt)\vec{X}_{[t=n\,dt]}
+Y_{\rm num}
\left(\begin{tabular}{c}
$d\xi_{[t=n\,dt]}$ \\ $d\xi^*_{[t=n\,dt]}$
\end{tabular}
\right)
\label{eq:recurX}
\end{equation}
with the initial condition $\vec{X}_{[t=0]}=0$. In this recursion
relation the friction matrix $\alpha_{\rm num}$ and the noise matrix
$Y_{\rm num}$ may differ
from $\alpha$ and $Y$ of the continuous
stochastic differential equation (\ref{eq:brown}) by terms linear
in $dt$ that remain to be determined in order to achieve an error scaling 
as $dt^2$.

As we have already mentioned $\vec{X}_{[t=n\,dt]}$ is a Gaussian vector
for any step $n$ of the iteration so that its probability distribution
is characterised by the covariance matrix $C_{ij}^{(n)}=\langle X_i X_j^*
\rangle$, with indices $i,j$ ranging from 1 to $2{\cal N}$. 
From (\ref{eq:recurX}) the covariance matrices
are shown to obey the recursion relation:
\begin{equation}
\label{eq:recur}
C^{(n+1)} = (1-\alpha_{\rm num}dt) C^{(n)} (1-\alpha_{\rm num}^{\dagger}dt)
+\frac{2dt}{dV}Y_{\rm num}Y_{\rm num}^\dagger.
\end{equation}
For a small enough time step $dt$ this matrix sequence converges to
a finite covariance matrix solving
\begin{equation}
\label{eq:condi}
C^{(\infty)} = (1-\alpha_{\rm num}dt) C^{(\infty)} (1-\alpha_{\rm num}^{\dagger} dt)
+\frac{2dt}{dV}Y_{\rm num}Y_{\rm num}^\dagger.
\end{equation}
We now try to choose the friction matrix and the noise matrix in order to
minimise the deviation of $C^{(\infty)}$ from the desired value, which is
the covariance
matrix of the exact distribution (\ref{eq:Pperp}), equal to $(2M\, dV)^{-1}$.
We look for $\alpha_{\rm num}$ and $Y_{\rm num}$
differing from the theoretical values (\ref{eq:mieux_alpha},\ref{eq:mieux_Y}) by
terms linear in $dt$, and leading to a covariance matrix different
from the theoretical one by terms quadratic in $dt$:
\begin{eqnarray}
\alpha_{\rm num } &=& 2 M + \alpha_1 dt \\
Y_{\rm num } &=& 
\left(\begin{array}{cc}
{\cal Q} & 0 \\
0 & {\cal Q}^* \\
\end{array}
\right) + Y_1 dt \\
C^{(\infty)} &=& \frac{1}{2M\, dV} + O(dt^2).
\end{eqnarray}
Equation (\ref{eq:condi}) is satisfied up to order $dt$ irrespectively of
the choice of $\alpha_1$, $Y_1$.
Requiring that equation (\ref{eq:condi}) is satisfied up to order $dt^2$ leads
to the condition
\begin{equation}
-\alpha_1\frac{1}{4M}-\frac{1}{4M}\alpha_1+
Y_1
\left( \begin{tabular}{cc}
${\cal Q}$ & 0 \\
0 & ${\cal Q}^*$
\end{tabular} \right)
+
\left( \begin{tabular}{cc}
${\cal Q}$ & 0 \\
0 & ${\cal Q}^*$
\end{tabular} \right)
Y_1^\dagger + M =0.
\end{equation}
A particular solution of this equation is provided by $\alpha_1 =0$
and $Y_1=Y_1^\dagger = -M/2$. Our improved integration scheme is therefore
\begin{eqnarray}
\label{eq:choix1}
\alpha_{\rm num } &=& 2M \\
\label{eq:choix2}
Y_{\rm num} &=& 
\left(\begin{array}{cc}
{\cal Q} & 0 \\
0 & {\cal Q}^* \\
\end{array}
\right) - \frac{1}{2} Mdt.
\end{eqnarray}

The analysis of the recursion relation (\ref{eq:recur}) is easily performed
for our improved integration scheme (\ref{eq:choix1},\ref{eq:choix2})
since $\alpha_{\rm num}$, $\alpha_{\rm num}^\dagger$,  $Y_{\rm num}$ and
hence $C^{(n)}$   
are polynomials of $M$ and commute with $M$. As a consequence $C^{(\infty)}$
also commutes with $M$.

Let us first estimate the deviation of $C^{(\infty)}$ from the
exact covariance matrix $(2M\, dV)^{-1}$:
\begin{eqnarray}
C^{(\infty)} &=& \left[1-\left(1-\alpha_{\rm num } dt\right)^2\right]^{-1}
\frac{2 dt}{dV} Y_{\rm num} Y_{\rm num}^{\dagger} \\
&\simeq& \frac{1}{2 M\, dV} \left[1+ \frac{dt^2}{4}M^2 + O(dt^3)\right].
\end{eqnarray}
Because $M$ is bounded from above by unity we take in practice $dt=1/8$ so
that the error
is less than 0.5 percent.

Let us finally estimate the convergence time of the covariance matrices. The recursion
relation (\ref{eq:recur}) can be rewritten as
\begin{equation}
C^{(n+1)} - C^{(\infty)} = (1-\alpha_{\rm num}dt)^2\left[C^{(n)} - C^{(\infty)}\right]
\end{equation}
so that the relative deviation of $C^{(n)}$ from its asymptotic value evolves
as $(1-2M_{\rm min}dt)^{2n}$ where $M_{\rm min}$ is the smallest eigenvalue
of $M$, that can be evaluated along the lines of \cite{Kuhtai}. We choose the
number of time steps $n$ so that the relative deviation 
of $C^{(n)}$ from $C^{(\infty)}$ is less than 0.5 percent.

\section{Moments of $N_0$ of a harmonically trapped ideal Bose condensed 
gas}
\label{appendix:moments}
We explain how to calculate the approximate expressions
(\ref{eq:asympt}) for the moments of the number of condensed particles
for an ideal Bose gas
in an isotropic harmonic potential of frequency $\omega$
in the temperature regime
$k_B T\gg\hbar\omega$ and in the Bogoliubov approximation.
The calculation of the moments involves sums over
the excited harmonic levels, see (\ref{eq:sommes}).
By using the known degeneracy of the harmonic eigenstate
manifold of energy $n\hbar\omega$ above the ground state energy
the calculation reduces to the evaluation of sums of the form
\begin{equation}
S_{p,q}(\epsilon) = \sum_{n=1}^{\infty} \frac{n^p}{(\exp(n\epsilon)-1)^q}
\end{equation}
where $\epsilon=\hbar\omega/k_B T$ is tending to zero, and the exponents
$p$ and $q$ are positive integers.

\noindent{\sl First case:} $q-p>1$: In the limit $\epsilon\rightarrow 0$
the sum is dominated by the contribution of 
small values of $n$.
Replacing $\exp(\epsilon n)-1$ by its first order expression we obtain:
\begin{equation}
S_{p,q}(\epsilon) \simeq \frac{1}{\epsilon^q}\sum_{n=1}^{\infty}\frac{1}
{n^{q-p}} = \frac{1}{\epsilon^q} \zeta(q-p)
\end{equation}
where $\zeta(\alpha)=\sum_{n\geq 1} 1/n^\alpha$ is the Riemann Zeta
function.

\noindent{\sl Second case:} $q-p <1$: In the limit $\epsilon\rightarrow 0$
the contribution to the sum is dominated by large values of $n$.
We then replace  the discrete sum by an integral over $n$ from 
$1$ to $+\infty$. Taking as integration variable $u=\epsilon n$
we arrive at
\begin{equation}
S_{p,q}(\epsilon) \simeq 
\frac{1}{\epsilon^{p+1}}
\int_{\epsilon}^{+\infty} du\, \frac{u^p}{(\exp(u)-1)^q}.
\end{equation}
We can take the limit $\epsilon\rightarrow 0$ in the lower bound
of the integral since $q-p<1$:
\begin{equation}
S_{p,q}(\epsilon) \simeq \frac{1}{\epsilon^{p+1}} I_{p,q}.
\end{equation}
To calculate the resulting integral $I_{p,q}$
we expand the integrand in series of $\exp(-u)$ and integrate term
by term over $u$:
\begin{equation}
\label{eq:ipq}
I_{p,q}\equiv \int_{0}^{+\infty} du\, \frac{u^p}{(\exp(u)-1)^q}
=\sum_{k=0}^{\infty} \frac{p!}{(k+q)^{p+1}}
\frac{(k+q-1)!}{k!(q-1)!}
\end{equation}
which can be expressed in terms of the Riemann Zeta function, e.g.
$I_{2,2}=2(\zeta(2)-\zeta(3))$.

\noindent{\sl Third case:} $q-p=1$: In the limit $\epsilon\rightarrow 0$
both the small values of $n$ and the large values of $n$ contribute to the 
sum. We introduce a small parameter $\nu\ll 1$ that will be put to zero at
the end of the calculation. For the summation indices $n<\nu/\epsilon$ 
we keep a discrete sum and we approximate each term of the sum by
its first order expression in $\epsilon$, which is correct as
$n\epsilon <\nu\ll 1$. For the summation indices $n>\nu/\epsilon$ 
we replace the sum by an integral, which is correct in the limit $\epsilon
\rightarrow 0$ for a fixed $\nu$, since we then recognise a Riemann
sum
of a function with a converging integral. This leads to
\begin{equation}
S_{p,p+1} \simeq \frac{1}{\epsilon^{p+1}}\left[
\sum_{n=1}^{\nu/\epsilon}\frac{1}{n} + \int_\nu^{+\infty}
du\;\frac{u^p}{(\exp(u)-1)^{p+1}}\right].
\end{equation}
In the limit $\epsilon\rightarrow 0$ the discrete sum is approximated by
\begin{equation}
\label{eq:Euler}
\sum_{n=1}^{\nu/\epsilon}\frac{1}{n} \simeq \log(\nu/\epsilon) +\gamma
\end{equation}
where $\gamma$ is Euler's constant. In the integral we remove and add
$1/(\exp(u)-1)$ to the integrand in order to get a convergent
integrand which facilitates the calculation of
the $\nu\rightarrow 0$ limit. The integral of $1/(\exp(u)-1)$
can be calculated explicitly from the primitive $\log(1-\exp(-u))$
so that in the small $\nu$ limit
\begin{eqnarray}
\int_\nu^{+\infty} du\;\frac{u^p}{(\exp(u)-1)^{p+1}} &=&
\log\frac{1}{1-\exp(-\nu)}+
\int_\nu^{+\infty} du\; \left[\frac{u^p}{(\exp(u)-1)^{p+1}}
-\frac{1}{\exp(u)-1}\right]   \\
&\simeq& -\log\nu + J_p 
\end{eqnarray}
where 
\begin{equation}
J_p = \int_0^{+\infty} du\; \left[\frac{u^p}{(\exp(u)-1)^{p+1}}
-\frac{1}{\exp(u)-1}\right].
\end{equation}
The $-\log\nu $ term coming from the integral compensates
the $\log\nu$ term coming from the sum in (\ref{eq:Euler})
so that in the limit $\nu\rightarrow 0$ we get the $\nu$-independent
estimate
\begin{equation}
S_{p,p+1} \simeq \frac{1}{\epsilon^{p+1}} \left[-\log\epsilon + \gamma
+ J_p\right].
\end{equation}
The quantity $J_p$ for $p>0$ 
can be calculated from a recursion relation obtained in the following
way: we use the identity
\begin{equation}
\frac{u^p}{(\exp(u)-1)^{p+1}}
= -\frac{u^p}{(\exp(u)-1)^p} +u^p\frac{\exp(u)}{(\exp(u)-1)^{p+1}}.
\end{equation}
The first term of the above expression leads to an integral already
calculated in (\ref{eq:ipq}) and called $I_{p,p}$.
We then integrate the second term of the above expression by parts, 
taking the derivate of $u^p$ with respect to $u$. This finally leads to
\begin{equation}
J_p = J_{p-1} +\frac{1}{p} - I_{p,p}.
\end{equation}
We get in particular $J_1=1-\zeta(2)$ and
$J_2=3/2-3\zeta(2)+2\zeta(3)$.

Finally we collect the approximations for the $S_{p,q}$ relevant for
the calculation of the skewness of the number of condensed particles
$N_0$ in 1D, 2D, 3D:
\begin{equation}
\begin{array}{lll}
\displaystyle
S_{0,1} \simeq \frac{-\log\epsilon +\gamma}{\epsilon}\quad &
\displaystyle
S_{0,2} \simeq \frac{\zeta(2)}{\epsilon^2} &
\displaystyle
S_{0,3} \simeq \frac{\zeta(3)}{\epsilon^3}  \\
& & \\
\displaystyle
S_{1,1} \simeq \frac{\zeta(2)}{\epsilon^2} &
\displaystyle
S_{1,2} \simeq \frac{-\log(\epsilon)+\gamma+1-\zeta(2)}{\epsilon^2}
\quad &
\displaystyle
S_{1,3} \simeq \frac{\zeta(2)}{\epsilon^3} \\
& & \\
\displaystyle
S_{2,1} \simeq \frac{2\zeta(3)}{\epsilon^3}  &
\displaystyle
S_{2,2} \simeq \frac{2\zeta(2)-2\zeta(3)}{\epsilon^3}  &
\displaystyle
S_{2,3} \simeq \frac{-\log\epsilon+\gamma+J_2}{\epsilon^3} \\
\end{array}
\end{equation}

\section{Equations of the number conserving Bogoliubov approach}
\label{app:BogolCD}

In this appendix we give the equations of motion 
for the operator $\hat{\Lambda}$ and for  $\phi_\perp^{(2)}(r)$
from \cite{CastinDum}.  
The evolution equation for  $\hat{\Lambda}$ is:
\begin{equation}
i\hbar \partial_t \left(\begin{array}{c}
\hat{\Lambda}(r,t) \\ 
\hat{\Lambda}^\dagger(r,t) \end{array}\right) = {\cal L}(t)
\left(\begin{array}{c} \hat{\Lambda}(r,t) \\ 
\hat{\Lambda}^\dagger(r ,t)
\end{array}\right) 
\end{equation}
with ${\cal L}$ given by (\ref{eq:calL}).  
The evolution equation for  $\phi_\perp^{(2)}(r)$ is:
\begin{equation}
 \left( i \hbar {d \over dt} - {\cal L}(t) \right)
\left( \begin{array}{c}
\phi_\perp^{(2)}(t) \\ 
\phi_\perp^{(2)*}(t)
\end{array} \right)
= 
\left( \begin{array}{r}
{\cal Q}(t)S(t) \\
-{\cal Q}^*(t)S^*(t)
\end{array} \right)
\end{equation}
where 
\begin{eqnarray} \label{SOURCE}
S(r) &=& -g N
{| \, {\phi({r})} \, |}^2 \phi({r})
\langle 1+ \sum_s dV\, {\hat{\Lambda}}^\dagger(s) 
	{\hat{\Lambda}}(s)\rangle
\nonumber \\
&+& 2 g N \phi({r})
\langle{\hat{\Lambda}}^\dagger({r}){\hat{\Lambda}}({r})\rangle
+g N \phi^*({r})
\langle{\hat{\Lambda}}({r}){\hat{\Lambda}}({r})\rangle
\nonumber \\
&-&g N \sum_s dV\, {| \, {\phi(s)} \, |}^2
 \langle\left[{\hat{\Lambda}}^\dagger(s)\phi(s)+ 
{\hat{\Lambda}}(s)\phi^*(s)\right] 
{\hat{\Lambda}}({r})\rangle \,.
\label{eq:S}
\end{eqnarray}

\section{Truncated Wigner approach in the Bogoliubov regime}
\label{app:Bogol}
In this appendix we demonstrate the equivalences
(\ref{eq:id_psi0}-\ref{eq:id_phi2}). For convenience we change in this
appendix the
phase reference of the field $\psi$ which now evolves according to
\begin{equation}
i\hbar \partial_t \psi = \left[p^2/2m
+U(r,t) + g |\psi|^2-\mu \right]\psi
\label{eq:GPEmod}
\end{equation}
where $\mu$ is the chemical potential in the time-independent
Gross-Pitaevskii equation for the condensate wavefunction (\ref{eq:tigpe}).

\begin{enumerate}
\item{\bf Identification of the pure condensate wavefunction}

At $t=0$ equation (\ref{eq:id_psi0}) is satisfied.
By keeping only terms of order
$\sqrt{N}$ in (\ref{eq:GPEmod}), in the limit $(\ref{eq:limit})$, we obtain 
\begin{equation}
i \hbar \partial_t \psi^{(0)} = (h_0 +g|\psi^{(0)}|^2-\mu) \psi^{(0)}
\end{equation}
where $h_0$ is the one-body part of the Hamiltonian.
This shows that (\ref{eq:id_psi0}) holds at all times.

\item{\bf ``Orthogonal-orthogonal'' contribution}

We wish to prove (\ref{eq:lin}).
To this aim we expand $\hat{\Lambda}$ and $\psi^{(1)}_\perp$ over the
Bogoliubov modes:
\begin{eqnarray}
\hat{\Lambda}=\sum_{k} \hat{ b}_k u_k + \hat{ b}_k^\dagger v_k^* 
\label{eq:explambda}\\
\psi^{(1)}_\perp=\sum_{k} { b}_k u_k + { b}_k^* v_k^* 
\label{eq:exppsi1}
\end{eqnarray}
At $t=0$ the same mode functions $u_k$ and $v_k^*$ appear in the 
expansions of  $\hat{\Lambda}$ and $\psi^{(1)}_\perp$. We wish to show that 
(\ref{eq:explambda}-\ref{eq:exppsi1}) hold at any time, or equivalently
that $\hat{\Lambda}$ and $\psi^{(1)}_\perp$ have the same equations of motion.
If we keep only terms of order $O(1)$ in (\ref{eq:GPEmod}) we get
\begin{equation}
i\hbar \partial_t \left(\begin{array}{c}
\psi^{(1)} \\ \psi^{(1)*} \end{array}\right) = {\cal L}_{GP} 
\left(\begin{array}{l} 
\psi^{(1)} \\ \psi^{(1)*} \end{array}\right)
\end{equation}
where ${\cal L}_{GP}$ is the usual Bogoliubov operator obtained from
(\ref{eq:calL}) by eliminating all the projectors.
By using the fact that 
\begin{equation}
\left( \begin{array}{c} \psi^{(1)}_\perp \\ 
	\psi^{(1)*}_\perp \end{array} \right) = 
\left( \begin{array}{cc} {\cal Q} & 0 \\ 0 & {\cal Q}^* \end{array}\right) 
\left( \begin{array}{c} \psi^{(1)} \\ \psi^{(1)*} \end{array}\right)  
\end{equation}
and
\begin{equation}
\left( \begin{array}{c} \xi^{(1)}  \phi \\ 
                        \xi^{(1)*} \phi^* \end{array}\right) = 
\left( \begin{array}{cc} {\cal P} & 0 \\ 0 & {\cal P}^* \end{array}\right) 
\left( \begin{array}{c} \psi^{(1)} \\ \psi^{(1)*} \end{array}\right)  
\end{equation}
with the matrices
\begin{equation}
{\cal P}_{r,s}= dV\,\phi(r) \phi^*(s) \hspace{1cm}
{\cal Q}_{r,s}=\delta_{r,s}-dV\phi(r)\phi^*(s)
\end{equation}
we get
\begin{eqnarray}
i\hbar \partial_t 
\left( \begin{array}{c} \psi^{(1)}_\perp \\ 
\psi^{(1)*}_\perp \end{array}\right) &=& {\cal L}
\left( \begin{array}{c} \psi^{(1)}_\perp \\ 
	\psi^{(1)*}_\perp \end{array}\right) +
(\xi^{(1)}+\xi^{(1)*}) 
\left( \begin{array}{cc} {\cal Q} & 0 \\ 0 & {\cal Q}^* \end{array}\right) 
\left( \begin{array}{c} gN |\phi|^2 \phi \\ 
                        -gN |\phi|^2 \phi^* \end{array}\right) 
\label{eq:eqpsi1} \\ 
i\hbar \frac{d}{dt} \xi^{(1)} &=&
        dV\sum_r  gN |\phi(r)|^2 [ \phi^{*}(r) \psi^{(1)}(r) +
        \psi^{(1)*}(r) \phi(r) ].
\label{eq:eqxi1}
\end{eqnarray}
The fact that the derivative of $\xi^{(1)}$ is purely imaginary and the
initial condition $\xi^{(1)}=0$ guarantee that $(\xi^{(1)}+\xi^{(1)*})=0$ 
for all times, which proves
that $\hat{\Lambda}$ and $\psi^{(1)}_\perp$ have the same equations of motion. 
At all times we then have
\begin{equation}
\langle \hat{\Lambda}^\dagger(s) \hat{\Lambda}(r) \rangle = \sum_k 
	u_k(r) u_k^*(s) \langle \hat{b}_k^\dagger \hat{b}_k  \rangle +
	v_k^*(r) v_k(s) \langle \hat{b}_k \hat{b}_k^\dagger  \rangle 
\end{equation}
and
\begin{equation}
\langle \psi^{(1)*}_\perp(s) \psi^{(1)}_\perp(r) \rangle = 
	\langle \hat{\Lambda}^\dagger(s) \hat{\Lambda}(r) \rangle +
\frac{1}{2}     \sum_k u_k(r) u_k^*(s) -  v_k^*(r) v_k(s) 
\end{equation}
where the amplitudes $b_k$ are time-independent
and the $u_k,v_k$ are time-dependent modes evolving according to
\begin{equation}
i\hbar \partial_t 
\left( \begin{array}{c} u_k \\ 
v_k \end{array}\right) = {\cal L}
\left( \begin{array}{c} u_k \\ 
        v_k \end{array}\right)\,.
\end{equation}

By using the decomposition of unity, equation (61) of reference
\cite{CastinDum}:
\begin{equation}
\sum_k u_k(r) u_k^*(s) -  v_k^*(r) v_k(s) = \frac{1}{dV}{\cal Q}_{r,s}
\end{equation}
we prove (\ref{eq:lin}).

\item {\bf ``Parallel-parallel'' contribution}

We wish to prove (\ref{eq:id_deltaN}).
We use the fact that
$\langle dV\sum_r |\psi(r)|^2\rangle$ is a constant of motion
order by order in $1/\sqrt{N}$.
To order $\sqrt{N}$ we get
\begin{equation}
\frac{d}{dt} N=0
\end{equation}
To order $N^0$ we get
\begin{equation}
\frac{d}{dt} \langle \xi^{(1)} + \xi^{(1)*} \rangle =0
\end{equation}
which we verified directly in (\ref{eq:eqxi1}).
To order $1/\sqrt{N}$ we get
\begin{equation}
\frac{d}{dt} \left[ \langle \xi^{(2)}+\xi^{(2)*} \rangle + 
\langle |(\xi^{(1)}|^2 \rangle + \langle dV\, \sum_r |\psi^{(1)}_\perp(r)|^2
\rangle \right] =0\,.
\end{equation}
Using (\ref{eq:lin}) we then obtain
\begin{equation}
\langle \xi^{(2)}+\xi^{(2)*} \rangle +
\langle |(\xi^{(1)}|^2 \rangle + \langle \delta \hat{N} \rangle 
+ \frac{{\cal N}-1}{2} = \mbox{constant}\,. \label{eq:aux1}
\end{equation}
At $t=0$ from (\ref{eq:ordreun}), (\ref{eq:ordredeux}) we deduce
\begin{equation}
\mbox{constant}  = \frac{{\cal N}-1}{2} 
\end{equation}
so that at any time
\begin{equation}
\langle \xi^{(2)}+\xi^{(2)*} \rangle +
\langle |(\xi^{(1)}|^2 \rangle = -\langle \delta \hat{N} \rangle\,.
\label{eq:manque_un_demi}
\end{equation}
Note that without the approximation in \cite{PRL} 
we would have at $t=0$ $\mbox{constant}  = \frac{\cal N}{2}$ and
as a consequence 
$\langle \xi^{(2)}+\xi^{(2)*} \rangle +
\langle |(\xi^{(1)}|^2 \rangle = -\langle \delta \hat{N} \rangle
+ \frac{1}{2}$.
The contribution of the 1/2 compensates exactly the term 
$-\frac{1}{2} \phi^*(s) \phi(r)$ in (\ref{eq:rho1wig}). We neglect here
this contribution.

\item{\bf Term ``parallel-orthogonal''}

The last step consists in proving (\ref{eq:id_phi2}). 
We first remark that at $t=0$ $\langle \psi^{(1)}_\perp \rangle=0$, and
for linearity reasons $\langle \psi^{(1)}_\perp \rangle=0$ at all times.
At $t=0$ (\ref{eq:id_phi2}) is satisfied by construction.
We then have to deduce the
equation of motion for
\begin{equation}
\langle \chi \rangle \equiv 
	\langle \xi^{(1)*} \psi^{(1)}_\perp + \psi^{(2)}_\perp 
	\rangle  
\end{equation}
and show that it coincides with the equation of motion for $\phi_\perp^{(2)}$.
By keeping only terms of order $1/\sqrt{N}$ in (\ref{eq:GPEmod}) we get
\begin{equation}
i\hbar \partial_t \left(\begin{array}{c}
\psi^{(2)} \\ \psi^{(2)*} \end{array}\right) = {\cal L}_{GP}
\left(\begin{array}{l}
\psi^{(2)} \\ \psi^{(2)*} \end{array}\right) +
 \left(\begin{array}{r}
gN [\phi^{*} \psi^{(1)2} + 2 \phi |\psi^{(1)}|^2] \\
-gN [ \phi\, \psi^{(1)*2} + 
        2 \phi^{*} |\psi^{(1)}|^2] \end{array}\right)\,.
\end{equation}
With a calculation analogous to the one we performed to obtain
the derivative of $(\psi^{(1)}_\perp,\psi^{(1)*}_\perp)$,  using
(\ref{eq:aux1}) to eliminate $\xi^{(2)}$ and replacing $\psi^{(1)}$ 
by $\xi^{(1)}\phi+\psi^{(1)}_\perp$,
we obtain:
\begin{eqnarray}
i\hbar \partial_t \left(\begin{array}{c}
\psi^{(2)}_\perp \\ \psi^{(2)*}_\perp \end{array}\right) &=& {\cal L}
\left(\begin{array}{c}
\psi^{(2)}_\perp \\ \psi^{(2)*}_\perp \end{array}\right) - 
\langle \delta \hat{N} \rangle
\left(\begin{array}{c}
  g N {\cal Q} |\phi|^2 \phi \\ -  g N {\cal Q}^* |\phi|^2 \phi^*  \end{array}\right) \\
&+& \left(\begin{array}{c}
g N {\cal Q} [2 |\psi^{(1)}_\perp|^2 \phi + 
	2 \xi^{(1)} \phi^2 \psi^{(1)*}_\perp +
        \phi^{*} (\psi^{(1)}_\perp)^2] \\
- g N {\cal Q}^* [2 |\psi^{(1)}_\perp|^2 \phi + 
	2 \xi^{(1)} \phi^2 \psi^{(1)*}_\perp +
        \phi^{*} (\psi^{(1)}_\perp)^2]^* 
        \end{array}\right)\,.
\end{eqnarray}
In particular, we find that the terms involving $|\xi^{(1)}|^2$ disappear because
$(\xi^{(1)})^2=-|\xi^{(1)}|^2$.
By using (\ref{eq:eqpsi1}) and (\ref{eq:eqxi1}) we can calculate the derivative 
of $\langle \chi \rangle$:
\begin{equation}
i\hbar \partial_t \left(\begin{array}{c}
\langle \chi \rangle \\ \langle \chi^* \rangle \end{array}\right) =
{\cal L} \left(\begin{array}{c}
\langle \chi \rangle \\ \langle \chi^* \rangle \end{array}\right) +
\left(\begin{array}{c}
{\cal Q} \; R \\ -{\cal Q}^* \; R^* \end{array}\right)
\label{eq:eqchi}
\end{equation}
with
\begin{eqnarray}
R(r)&=&-\langle \delta \hat{N} \rangle gN |\phi(r)|^2 \phi(r) +
2gN \phi(r) [\langle \hat{\Lambda}^\dagger \hat{\Lambda} \rangle -
		\frac{1}{2} |\phi(r)|^2] \nonumber \\
&+& gN\phi^* \langle \hat{\Lambda}^2 \rangle -
gN \left\{
\frac{1}{2}\phi(r)|\phi(r)|^2+
dV \sum_s \; |\phi(s)|^2
\langle [\hat{\Lambda}^\dagger(s)\phi(s) +\phi^*(s)\hat{\Lambda}(s) ]
	\hat{\Lambda}(r) \rangle
         \right\}
\label{eq:R}
\end{eqnarray}
which is identical to (\ref{eq:S}), except for the contribution of
the term $1/2$ neglected in \cite{PRL} as discussed after
(\ref{eq:manque_un_demi}).
In order to obtain (\ref{eq:R}) we used the identity (\ref{eq:lin})
and the fact that all terms proportional to $\phi(r)$ are killed by 
the projector ${\cal Q}$ in (\ref{eq:eqchi}). Summarising,
(\ref{eq:eqchi}) and
(\ref{eq:R}) together with  $\langle \psi^{(1)}_\perp \rangle=0$ prove
(\ref{eq:id_phi2}).
\end{enumerate}

\section{Equation for the noncondensed field in the Wigner approach}
\label{app:eqlambda_ex}
In the truncated Wigner approach, we define the field $\Lambda_{\rm ex}(r)=a_{\phi}^*\psi_\perp(r)/\sqrt{N}$
where $\phi$ is at this stage an arbitrary wave function normalised to unity, 
$\psi_\perp$ is the component of $\psi$ orthogonal to $\phi$, and $a_\phi$ is the coefficient of $\psi$
along $\phi$. When $\psi$ solves the time-dependent Gross-Pitaevskii equation,
the equation of motion for $\Lambda_{\rm ex}$ is given by:
\begin{eqnarray}
i\hbar \frac{d \Lambda_{\rm ex}}{dt}=
{1 \over \sqrt{N}} i \hbar {d \over dt}  \left(a_{\phi}^*
{{\psi_\perp}}({r}) \right)
= dV\sum_s \sum_{k=0}^{4} \frac{R_k({r},{s})}{N^{(k-1)/2}}
\end{eqnarray}
where we have collected the terms of the same power in $\Lambda_{\rm ex}$:
\begin{eqnarray}
R_0(r,s) &=& \frac{N_\phi}{N} {\frac{{\cal Q}_{r,s}}{dV} 
{ [ -i \hbar\partial_t + {h_0} +g N_\phi 
	| \, {\phi({s})} \, |^2] \phi({s})         } }  \nonumber \\
R_1(r,s) &=& \frac{{\cal Q}_{r,s}}{dV} 
\left[{h_0} + 2 g N_\phi {| \,
{\phi_{}(s)} \, |}^2\right] \Lambda_{\rm ex}(s) -
\phi_{}(r) (i \hbar\partial_t \phi_{}^*(s)) \Lambda_{\rm ex}(s) \nonumber \\ 
&+& {\frac{{\cal Q}_{r,s}}{dV} { g N_\phi \phi_{}^2(s) 
\Lambda^*_{\rm ex}(s)     } } - \Lambda_{\rm ex}(r)
\phi_{}^*(s) ( -i \hbar\partial_t +{h_0} +g N_\phi {| \, {\phi_{}(s)} \, |}^2)
  \phi_{}(s)
    \nonumber \\
R_2(r,s)&=&  - \frac{N}{N_\phi}
        \Lambda_{\rm ex}^*(s) \Lambda_{\rm ex}(r)
( -i \hbar\partial_t +{h_0} +2 g N_\phi {| \, {\phi_{}(s)} \, |}^2) \phi_{}(s)
 \nonumber \\
&+& g N {\frac{{\cal Q}_{r,s}}{dV} {} }\left[ \Lambda_{\rm ex}^2(s) \phi_{}^*(s)
+ 2 \Lambda_{\rm ex}^*(s) \Lambda_{\rm ex}(s) \phi(s)
       \right] \nonumber \\
&-& g N \phi_{}^*(s) {| \, {\phi_{}(s)} \, |}^2
   \Lambda_{\rm ex}(s) \Lambda_{\rm ex}(r) 
\nonumber \\
R_3(r,s) &=&   
 g N \frac{N}{N_\phi} \left[ {\frac{{\cal Q}_{r,s}}{dV}
\Lambda_{\rm ex}^*(s) \Lambda_{\rm ex}^2(s) }
- \Lambda_{\rm ex}^{*2}(s) \Lambda_{\rm ex}(r) \phi^2(s)\right]
  \nonumber \\
&-&2 g N \frac{N}{N_\phi} {| \, {\phi_{}(s)} \, |}^2 
\Lambda_{\rm ex}^*(s) \Lambda_{\rm ex}(s) \Lambda_{\rm ex}(r) 
 \nonumber \\
R_4(r,s)&=& 
-gN \left(\frac{N}{N_\phi}\right)^2
\Lambda_{\rm ex}^{*2}(s) \Lambda_{\rm ex}(s) \Lambda_{\rm ex}(r)\phi(s)
\label{eq:lambdaTW}
\end{eqnarray}
where $N_\phi=a_\phi^* a_\phi$,
$h_0=p^2/2m + U(r,t)$ is the one-body part of the Hamiltonian
and ${\cal Q}_{r,s}=\delta_{r,s}-dV\phi(r)\phi^*(s)$
projects orthogonally to $\phi$.
In the case of a uniform wavefunction
$\phi(r)=1/L^{3/2}$ we have the
following simplifications:
(i) $\partial_t\phi$ is equal to zero,
(ii) the constant terms like $|\phi(s)|^2\phi(s)$ are killed by the
 projectors, 
(iii) for terms having a vanishing spatial sum,
$\frac{{\cal Q}_{r,s}}{dV}$ can be replaced by
$\delta_{r,s}$, (iv) the sum over $s$ of $\psi_\perp(s)$ and
therefore of $\Lambda_{\rm ex}(s)$ is zero.
For this value of $\phi$, $\Lambda_{\rm ex}$ coincides with $\Lambda_{\rm static}$
defined in (\ref{eq:classLambda}) and $N_\phi$ is equal to $N_0$ 
of equation (\ref{eq:simpleLambda}).


\begin{thebibliography}{99}
\bibitem{Drummond} M. J. Steel, M. K. Olsen, L. I. Plimak, P. D. Drummond,
S. M. Tan, M. J. Collett, D. F. Walls, and R. Graham, Phys. Rev. A {\bf 58}, 4824
(1998).
\bibitem{wqo}D. F. Walls and G. J. Milburn, {\it Quantum Optics}, 
Springer-Verlag, Berlin (1994).
\bibitem{PRL} A. Sinatra, C. Lobo and Y. Castin, Phys. Rev. Lett.
{\bf 87}, 210404 (2001).
\bibitem{Kuhtai} A. Sinatra, Y. Castin and C. Lobo, 
Jour. of Mod. Opt. {\bf 47}, 2629-2644 (2000).
\bibitem{CastinDum}Y. Castin and R. Dum, Phys. Rev. A {\bf 57} 3008-3021 (1998).
\bibitem{Gardiner} C. Gardiner, Phys. Rev. A {\bf 56} 1414-1423 (1997).
\bibitem{Iacopo} About the solution of the exact problem see e.g.
I. Carusotto, Y. Castin, and J. Dalibard, Phys. Rev. A {\bf 63}, 023606 (2001) and
L. I. Plimak, M. K. Olsen, M. Fleischhauer and M. J. Collett, 
Europhys. Lett. {\bf 56}, 372-378 (2001).
\bibitem{terme_omis}
Actually the equation for $\psi$ obtained in the truncated Wigner approach
differs from the usual Gross-Pitaevskii equation by a term involving a sum
over all modes $\phi_{k}(r)$ of the field.
In this paper we restrict ourselves to the case where $\sum_k \phi_k(r)
\phi_k^*(r)$ is a constant independent of the position
$r$. In this case the mean field term in the truncated Wigner
point of view differs from the one $g|\psi|^2$ of the usual Gross-Pitaevskii
equation by a constant term involving the number 
of modes, see \cite{Drummond} and our section \ref{sec:basics} below.
\bibitem{Kagan} Yu. Kagan and B. Svistunov, Phys. Rev. Lett. {\bf 79} 3331
(1997) and references therein.
\bibitem{Sachdev} K. Damle, S. Majumdar, and S. Sachdev,
Phys. Rev. A {\bf 54}, 5037 (1996).
\bibitem{Davis} M.J. Davis, S.A. Morgan and K. Burnett,
Phys. Rev. Lett. {\bf 87}, 160402 (2001).
\bibitem{Polonais} K. G\'oral, M. Gajda, K. Rz\c{a}\.zewski,
Opt. Express {\bf 8}, 82 (2001).
\bibitem{Burnett} M.J. Davis, S.A. Morgan and K. Burnett, cond-mat/0201571.
\bibitem{qnoise} C. Gardiner, {\em  Quantum Noise}, Springer-Verlag Berlin 
Heidelberg (1991), chapter 4 \lq\lq Phase space methods".
\bibitem{big_enough} 
In the case of a harmonically trapped gas
we assume that the box is large enough so that the atomic density
is small close to the boundaries.
\bibitem{Houches}Y. Castin, lectures in:
Coherent Atomic Matter Waves, Les Houches Summer School 
Session LXXII in 1999, edited by R. Kaiser, C. Westbrook, 
and F. David (Springer, New York, (2001), and cond-mat/0105058.
\bibitem{NumRecGauss} William H. Press, Brian P. Flannery,
Saul A. Teukolsky, William T. Vetterling, {\it Numerical Recipes}, \S 7.2,
Cambridge University Press (1986).
\bibitem{Wilkens} M. Wilkens, C. Weiss,  Opt. Expr {\bf 1}, 272 (1997).
\bibitem{Stringari} S. Giorgini, L. P. Pitaevskii, S. Stringari, Phys. Rev.
Lett. {\bf 80}, 5040, (1998). 
\bibitem{Scully} V.V. Kocharovsky, Vl. V. Kocharovsky, and Marlan O. Scully,
 Phys. Rev. Lett. {\bf 84}, 2306 (2000). 
\bibitem{bizarre} We have included not only the leading terms
but also higher order terms in the denominator of the 3D formula
since we have observed numerically that this dramatically improves
the accuracy of the formula for moderately high values
of $k_B T/\hbar\omega$.
\bibitem{future} see I. Carusotto and Y. Castin (in preparation).
\bibitem{PENROSE} exact in the sense of
O. Penrose and L. Onsager, Phys. Rev. {\bf 104}, 576 (1956).
\bibitem{enfin} We find at last the physical interpretation of a component of
$\phi^{(2)}$ that was unexplained in \cite{CastinDum}.
\bibitem{depletion} Y. Castin and R. Dum, Phys. Rev. Lett. {\bf 79}, 3553
(1997).
\bibitem{Bragg_MIT1} D. Stamper-Kurn, A. Chikkatur, A. G\"orlitz, S. Inouye,
S. Gupta, D. Pritchard, W. Ketterle, Phys. Rev. Lett. {\bf 82},
4569 (1999).
\bibitem{Bragg_MIT2} W. Ketterle, Spinor Condensates and Light Scattering
from Bose-Einstein Condensates, in Les Houches Summer
School 1999, Session LXXII, eds. R. Kaiser and C. Westbrook
(cond-mat/0005001)
\bibitem{singlereal} In the simulation we observe damping in each single 
realisation, which indicates that this damping is a true relaxation
phenomenon
and not a collapse due to dephasing among different stochastic
realisations.
This was not the case in the 1D model of \cite{PRL}.
\bibitem{quantique_aussi}
We note that exactly the same procedure from the beginning of this
subsection can be followed in the quantum treatment.
The evolution equation for $\hat{\Lambda}_{\rm ex}$ can in fact be obtained
from (\ref{eq:lambdaTW}) by putting hats on $a_\phi$ and $\psi_\perp$ and
by changing stars into daggers (see equation (A3) of \cite{CastinDum}).
\bibitem{intervalle} For the parameters of figure (\ref{fig:landau})
we perform a linear fit
of the perturbative prediction for $|\langle b_0(t)\rangle|$
on the time interval $(0.0075,0.015)$ in units of $mL^2/\hbar$ for
a grid size $48^3$. The slope is $-0.328$ with a linear correlation coefficient
$-0.99998$. The results are the same for a grid size $64^3$.
\bibitem{gamma_aussi} In the perturbative calculation for the figure
we have actually included the effect of the interaction between Bogoliubov
modes which provides damping
also in the excitation phase, simply by adding a term 
$-i\hbar\gamma_{\mbox{\scriptsize perturb}}b_0$ 
and $-i\hbar\gamma_{\mbox{\scriptsize perturb}}b_{-0}$ 
to the right hand side of
(\ref{eq:lin1}) and of (\ref{eq:lin2}) respectively. 
This damping term is not totally
negligible indeed since $\gamma_{\mbox{\scriptsize perturb}}t_{\rm exc}
\simeq 0.1$.
\bibitem{Vincent} Vincent Liu, Phys. Rev. Lett.  {\bf 79}, 4056 (1997).
\bibitem{Stringari_Pitaevskii}  L. P. Pitaevskii, and S. Stringari,
Phys. Lett. A {\bf  235}, 398 (1997).
\bibitem{Shlyapnikov}  P. O. Fedichev and G. V. Shlyapnikov, Phys. Rev. A {\bf
58}, 3146 (1998).
\bibitem{check_ener} 
We have checked with the particular example ${\cal N}
=22^3$, $N=5\times 10^4$, $k_B T= 3\mu$ and $\mu=500 \hbar^2/mL^2$,
that the mean Bogoliubov energy (after averaging
over 100 realisations) is conserved during the course of the time
evolution
at the 2\% level.
\bibitem{Modugno}  Michele Modugno, Ludovic Pricoupenko, and Yvan
Castin, cond-mat/0203597.
\bibitem{pourquoi} We have used the inequality $x/\tanh x < 1+x^2/3$.
\bibitem{Stoof} H. T. C. Stoof, J. Low Temp. Phys. {\bf 114}, 1 (1999).
\bibitem{Fudge} C. W. Gardiner, J. R. Anglin, and T. I. A. Fudge,
cond-mat/0112129.
\bibitem{NumRecCheb} Ibidem \cite{NumRecGauss},
\S 5.6.
\end{thebibliography}
\end{document}